\documentclass[reprint,
 amsmath,amssymb,
 aps,
 prstab,
 floatfix,
]{revtex4-2}

\usepackage{CJKutf8} % for name typesetting, somehow not working
\usepackage{graphicx}% Include figure files
\usepackage{dcolumn}% Align table columns on decimal point
\usepackage{bm}% bold math
\usepackage{siunitx}
\usepackage{algorithm}
\usepackage{algpseudocode}

\usepackage{cleveref}

\DeclareMathOperator{\argmax}{arg\,max}

\begin{document}
\begin{CJK}{UTF8}{gbsn}
%\begin{CJK*}{UTF8}{}
%\preprint{APS/123-QED}

\title{Bayesian Optimization of the Beam Injection Process into a Storage Ring}

\author{Chenran Xu 
(徐晨冉)
}
 \email{chenran.xu@kit.edu}
\author{Tobias Boltz}
 \thanks{Present address: SLAC, Menlo Park, USA}
\author{Akira Mochihashi}
\author{Andrea {Santamaria Garcia}}
\author{Marcel Schuh}
\author{Anke-Susanne M\"uller}
\affiliation{%
    Karlsruhe Institute of Technology, Kaiserstraße 12, 76131 Karlsruhe, Germany
}

% \date{\today}
%\date{February 14, 2022}

\begin{abstract}

We have evaluated the data-efficient Bayesian optimization method for the specific task of injection tuning in a circular accelerator.
In this paper, we describe the implementation of this method at the Karlsruhe Research Accelerator with up to nine tuning parameters, including the determination of the associated hyperparameters. 
We show that the Bayesian optimization method outperforms manual tuning and the commonly used Nelder-Mead optimization algorithm both in simulation and experiment.
The algorithm was also successfully used to ease the commissioning phase after the installation of new injection magnets and is regularly used during accelerator operations. We demonstrate that the introduction of context variables that include intra-bunch scattering effects, such as the Touschek effect, further improves the control and robustness of the injection process.
%We look forward to further extending this method by including safety constraints and applying it for other tuning tasks.

\end{abstract}

\maketitle
%\clearpage\end{CJK*}
\end{CJK}
%\tableofcontents
% \linenumbers

\section{\label{sec:introduction}Introduction}
Particle accelerators contribute to major discoveries in particle physics and are used to generate synchrotron radiation for photon science applications~\cite{Bruendermann2012}. 
% To enable this advanced level of performance, it is necessary to provide the best particle beam quality possible. 
They consist of a large and varied number of components, from magnets to steer, focus, or quickly redirect the particle beam, to radio frequency (RF) sources to accelerate it. 
This leads to a large number of possible parameter combinations for tuning, which is a complex and time consuming task. 
In addition, the beam dynamics are influenced by many non-linearly correlated parameters and are subject to physical phenomena such as the Touschek effect~\cite{Bernardini1963}. 
Although advanced control systems are employed, the fine tuning of beam properties is and will be a challenge, especially for future large-scale and compact accelerators. 
While intervention is required for several accelerator operation tasks,  manual tuning may not find the global optimum, especially in the shortest possible time, and does not adapt well to short-term and long-term drifts of the accelerator condition due to the large number of correlated parameters that influence it.

In such a case, computer algorithms can be introduced to assist the operator and eventually automate the tuning process~\cite{Edelen2018a}. Metaheuristics such as the evolutionary algorithm~\cite{Vikhar2016} and particle swarm optimization~\cite{Kennedy1995} need a large number of evaluation steps to converge and are not suited for online tuning. Gradient-based methods like robust conjugate direction search~\cite{Huang2015, Huang2018} are successfully used for online accelerator tuning, but they are prone to get stuck in local optima and observation noise.
Another promising local optimization method is the model independent extremum seeking (ES) algorithm. Despite also being sensitive to local optima, ES can be used as a feedback control to track the optimum settings with respect to drifts of the accelerator components. It has been successfully applied to tune time-varying accelerator systems with a large number of parameters~\cite{Scheinker2018minimization,Scheinker2019model}.
As a method to globally optimize an unknown function with expensive evaluations, Bayesian Optimization (BO) is shown to perform well among other approaches~\cite{Jones2001}. With observed data,
BO builds a surrogate model of the unknown function using a Gaussian process (GP)~\cite{Rasmussen2006} and uses an acquisition function to guide the search efficiently. Recently, BO has been successfully applied at the LCLS ~\cite{McIntire2016, Duris2020} and the SwissFEL~\cite{Kirschner2019} for free-electron laser performance tuning, motivating the usage of Bayesian optimization for other accelerator tuning tasks~\cite{Hanuka2019, Roussel2020}.

The process of beam injection can be approximately viewed as a black-box function optimization problem, which makes BO a well-suited method for the injection tuning.
To investigate this specific optimization problem, we use the accelerator test facility Karlsruhe Research Accelerator (KARA), which is a 110-meter storage ring and part of the KIT Light Source. It produces synchrotron radiation for photon science users and accelerator physics experiments at a top energy of \SI{2.5}{GeV}. The electron storage ring is filled at \SI{500}{MeV} by a chain of pre-accelerators with a repetition rate of \SI{1}{Hz}. The injection tuning is performed manually, and initial studies have shown that BO is well-suited to optimize this task~\cite{Xu2020}.
Furthermore, it is known that the injection condition may vary from one day to another and operators need to re-tune the injection elements. This might result from a change in the environmental conditions such as the ambient temperature or the quality of the vacuum , variables that are either non-controllable or can only be changed slowly.
In this article, besides showing the implementation of BO at a storage ring with several examples and use cases, we also take into account the effect of the environment by generalizing the standard BO method to contextual Bayesian optimization (CBO)~\cite{Krause2011} to optimize the injection efficiency under the given context.

\section{\label{sec:bo_intro}Bayesian Optimization}
Bayesian optimization (BO) is a method designed to optimize a black-box function $f$ in a sample-efficient way. The BO algorithm used in this paper is shown in pseudo-code in Algorithm~\ref{alg:bayesopt}~\cite{Brochu2010}. It makes use of a statistical surrogate model of the objective function $f$, generally built with a Gaussian Process (GP), based on the belief that the objective function $f$ is drawn from some prior probability distribution $p(A)$. After initialization and observations $f(x)$, the posterior distribution $p(A|f(x))$ is built according to the Bayes' theorem
\begin{equation}
\label{eq:bayes_theorem}
    p(A|f(x)) \propto p(f(x)|A)p(A).
\end{equation}
The posterior distribution is further used to build an acquisition function $\alpha (x)$, which determines the next point for evaluation.

\begin{algorithm}[H]
\caption{Bayesian optimization algorithm}
\label{alg:bayesopt}
\begin{algorithmic}[1]
        \State  Define the prior for the GP.
        \State  Observe objective function $f$ on $n_{0}$ initial points to get initial data set $D_{0}$.

        \For{$t=1,2,...$}
        \State Build a GP using available data $D_{t-1}$.
        \State Find the next evaluation point
        \Statex \hskip1.5em $x_{t} = \argmax_{x} \alpha (x \; \vert \; D_{t-1})$.
        \State Observe $y_{t}$ at point $x_{t}$.
        \State Augment the dataset $D_{t} = D_{t-1} \cup {(x_{t},y_{t})}$.
        \EndFor{}
\end{algorithmic}
\end{algorithm}

\subsection{\label{sec:gp}Gaussian Process}

A Gaussian process (GP) is a generalization of the multivariate Gaussian distribution and is characterized by its mean and covariance functions $f(x) \sim GP (\mu(x),k(x,x^{'}))$.
In this study, we set the prior mean function to be $\mu (x)=0$, which is commonly used when the shape of the objective function is unknown.
The covariance function $k(\cdot,\cdot)$, also known as kernel, measures the similarity between data points. Based on the assumption that $f$ is continuous, the data points $x$ that are close to each other are expected to have similar output values to $f(x)$. In this study, we use the widely used radial basis function (RBF) as the kernel function
\begin{equation}
\label{eq:rbf_kernel}
    k_\text{RBF} (x,x^{'}) = \exp \left(- \frac{1}{2} \sum^{d}_{i=1} \left(\frac{x_i-x_{i}^{'}}{l_i} \right)^{2} \right),
\end{equation}
where $l_i$ represents the length-scale of the i-th input dimension. It roughly corresponds to the distance along one input axis, at which the two data points become uncorrelated and the function values can change significantly.

In order to emulate the noise present in the real observed signal, we explicitly incorporate stochastic noise by adding Gaussian distributed noise to the covariance function as diagonal terms. Thus, the kernel becomes
\begin{equation}
\label{eq:full_kernel}
    k (x_i,x_j) = \sigma^{2} k_\text{RBF} (x_i,x_j) +  \sigma_{\text{noise}}^{2} \delta_{ij}.
\end{equation}

The signal variance $\sigma^{2}$, the noise variance $\sigma_{\text{noise}}^{2}$, and the length-scales $l_i$ are referred to as the hyperparameters, which determine the behavior of the GP.
Since no dedicated optimization data is stored in the database, several measurements are performed to estimate the hyperparameter settings.

\begin{figure}[tb]
\centering
\includegraphics[width=\columnwidth]{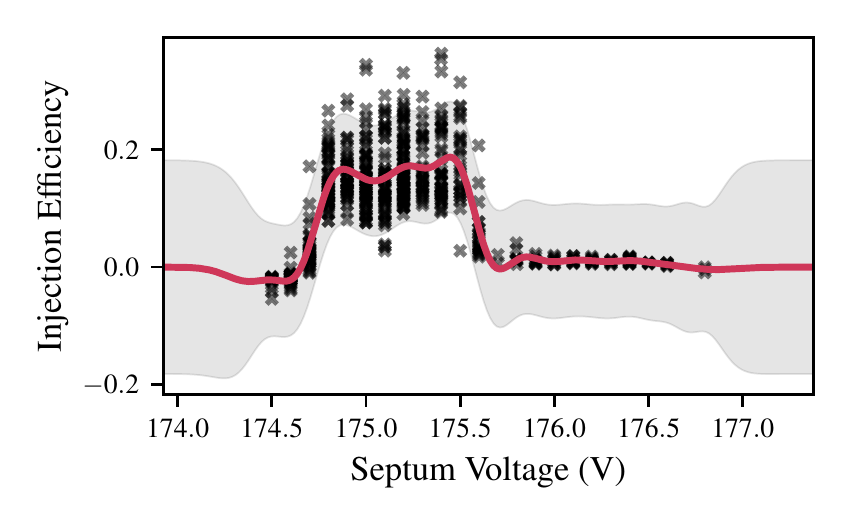}
\caption{\label{fig:param_scan} Example of a parameter scan measurement of the septum magnet voltage, where all other accelerator tuning parameters are kept constant during the scan. The GP hyperparameters $\{ l_i, \sigma^2 \}$ are fitted to the measured data points (crosses) via a log-likelihood fit. The solid line depicts the GP posterior mean and the shaded region shows the 95\% confidence level.}
\end{figure}

First, the noise $\sigma_\text{noise}$ is determined by fixing the accelerator settings and measuring the fluctuation of the objective function, i.e.\,the injection efficiency, as defined in Eq.~\ref{eq:objective_real}. By the assumption that the noise mostly comes from the statistical fluctuation between different injection shots, $\sigma_\text{noise}$ can be extracted as the standard deviation of a Gaussian fit.

Secondly, the length-scales $l_i$ and the signal variance $\sigma^2$ are estimated from one-dimensional parameter scans. Figure \ref{fig:param_scan} shows the scan result of the septum voltage, which corresponds to its magnetic field strength. We extract the GP hyperparameters with the maximum likelihood fit.

\subsection{\label{sec:acq}Acquisition Function}

With a GP model inferring the posterior distribution of the objective function, an acquisition function $\alpha$ is built to determine the next sample point to evaluate, so that the number of required physical observations is effectively reduced. Concretely, the objective $f$ is sampled at $\argmax_{x} \alpha(x|D)$ at each step, where $D$ is the observed dataset. We use two different acquisition functions in this study: the upper confidence bound (UCB) and the expected improvement (EI).
UCB explicitly controls the exploration-exploitation trade-off with a parameter $\kappa$
\begin{equation}
    \label{eq:UCB}
    \alpha_{\text{UCB}} (x) = \mu(x) + \kappa \sigma (x),
\end{equation}
where $\mu(x)$ and $\sigma(x)$ are the GP posterior mean and standard deviation. For high $\kappa$ values, the contribution of the variance term becomes large and points with high uncertainty are sampled, which leads to more exploration. On the contrary, small $\kappa$ values lead to more exploitation of observed peaks, as they have a higher posterior mean.

An empirical choice of the parameter is $\kappa=2$, which corresponds to the usual 95\% confidence bound for Gaussian distributed values. Nevertheless, $\kappa$ can also increase along with the evaluation steps to ensure that BO converges to the global optimum~\cite{Brochu2010,Srinivas2010}.

The expected improvement (EI) calculates the expected value of the improvement of a proposed point $x$ over the best observed value $f_{\text{best}}$~\cite{Jones1998}
\begin{equation}
\label{eq:EI}
\begin{split}
    \alpha_{\text{EI}} (x) & = \mathbb{E} [ \max (\mu (x)-(f_{\text{best}}+\xi),0)  ] \\
           & = (\mu(x)-(f_{\text{best}}+\xi)) \Phi (Z) + \sigma (x)\phi (Z)
\end{split}
\end{equation}
with the parameter $Z$ describing the normalized improvement
\begin{equation}
\label{eq:EI2}
\begin{split}
         Z  & = \frac{\mu(x)-(f_{\text{best}}+\xi)}{\sigma(x)},
\end{split}
\end{equation}
where $\phi$ is the probability distribution function and $\Phi$ is the cumulative distribution function of the standard normal distribution.
The exploration-exploitation trade-off in EI is determined by a positive parameter~$\xi$. In general, higher $\xi$ values lead to more exploration. In this study, we use the recommended value of $\xi=0.01$ according to Ref.~\cite{Lizotte2008}, which emphasizes exploitation.

%\subsection{\label{sec:context}Contextual Optimization}

%The basic BO algorithm can be extended to incorporate uncontrollable environmental variables, also called context~\cite{Krause2011}, where both the tuning parameters $s \in S$ and context variables $z\in Z$ are taken into account to build a joint GP over the tuning and context space $S \times Z$.

%The composite kernel can be constructed in different ways. For the RBF kernel which is used in this study, the kernels are often combined in the product form $k_\text{Product} = k_s \otimes k_z$. Thus, the resulting kernel remains a RBF kernel:
%\begin{equation}
%\label{eq:rbf_context}
%    k_\text{Product} (x,x^{'}) = \exp \left(- \frac{1}{2} \sum^{d_s+d_z}_{i=1} \left(\frac{x_i-x_{i}^{'}}{l_i} \right)^{2} \right),
%\end{equation}
%where $x=(s,z)$ is composed of $d_s$ tuning parameters and $d_z$ context variables.

%In addition to the normal BO (Algorithm \ref{alg:bayesopt}), the context variables $z_t$ are observed at each the optimization step. The acquisition function is then maximized over $S$ with the context dimensions fixed
%\begin{equation}
%    x_{t} = \argmax_{s\in S} \alpha ((s,z_{t}) \; \vert \; D_{t-1}).
%\end{equation}

\section{\label{sec:implementation_at_kara}Implementation at a storage ring}
In this section, we describe the implementation of the BO method to improve the injection efficiency into a storage ring, namely the storage ring of the KIT Light Source, the Karlsruhe Research Accelerator (KARA).

The electron storage ring is filled at \SI{500}{MeV} with a repetition rate of \SI{1}{Hz} by a chain of pre-accelerators, consisting of an electron gun, a microtron, and a booster synchrotron. Further details can be found in Ref.~\cite{Einfeld1998}.

\begin{figure}[tb]
\includegraphics[width=\columnwidth]{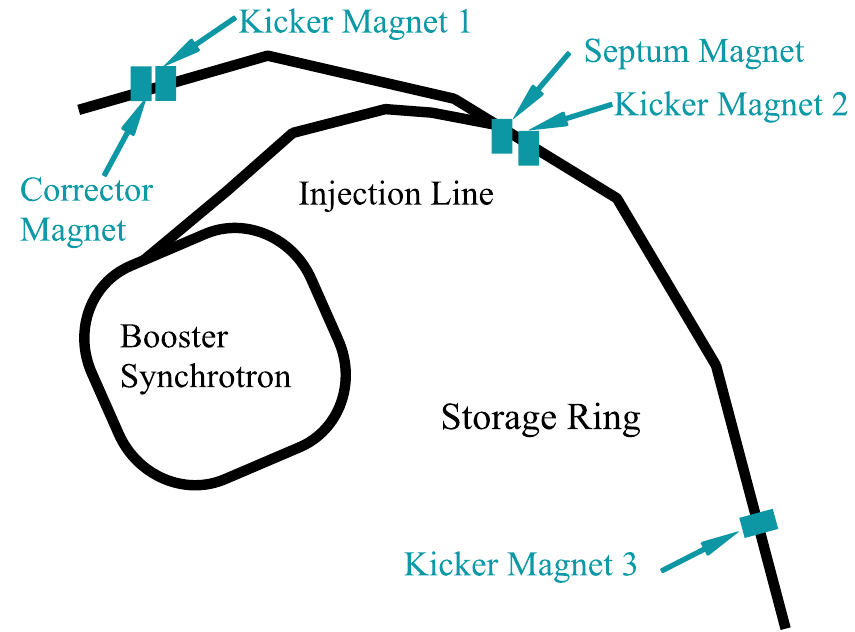}
\caption{\label{fig:injection_scheme} 
In the injection scheme, electron bunches are injected from a synchrotron via an injection line into the storage ring. The relative positions of the magnets and their location are illustrated as they are placed at the storage ring KARA at KIT. See Ref.~\cite{Einfeld1998} for further details. The tuning parameters of the magnets and RF cavities, not shown in this figure, are used as optimization parameters.}
\end{figure}

The storage ring employs a three-kicker injection scheme \cite{Turner:1994znj}, as shown in Fig.~\ref{fig:injection_scheme}, to create a closed orbit bump.
When the kicker magnets are powered they steer the closed orbit away from the nominal trajectory and near the septum magnet.
Kicker magnets are fast pulsed (in the order of \si{\micro s}) with a relatively low field strength, so they are combined with septa magnets that provide a stronger field.
The septum magnet deflects the beam into the aperture of the storage ring and provides a separation space between the circulating and injected beam.
The separation barrier (septum) is usually as thin as possible with minimum field leakage.
The injection bump orbit at KARA spans over one quadrant of the storage ring. There are several sextupoles in the injection bump which add non-linearities to the beam dynamics of the injection process. The stray field of the septum magnet and the potential energy mismatch between the booster and the storage ring also affect the settings required for a closed bump orbit. Moreover, due to the lack of non-destructive beam diagnostics in the transfer line except of one current monitor, the tuning is partly blind and relies mostly on the operators' experience.

In order to obtain a good injection rate, operators have to adjust multiple parameters like the magnet currents and radio frequency (RF) parameters. The manual tuning often ends up at a local optimum with lower injection rate, which eventually reduces the beamtime availability.
The tuning task after a shutdown period can be especially difficult, since previous accelerator settings can no longer be used and it costs valuable beamtime to recover the expected injection performance.
We show that the BO method outperforms the simplex methods and achieves a sample-efficient optimization.

We selected 9 tuning parameters for the BO based on their influence on the injection process and from prior knowledge of experienced operators. The parameters are: the main radio frequency (RF), the injection septum magnet strength, strength of one horizontal corrector magnet, and the strength and timing of the three kicker magnets. 
The strength and timing of the three kickers defines the shape of the injection bump orbit. The strength of the septum magnet changes the angle of the beam injected into the storage ring. The corrector magnet chosen here is located directly before the first kicker magnet and controls the angle of the incoming beam to the injection bump. Finally, the RF defines the orbit length and by changing it slightly the closed-orbit condition is modified. 
The relative positions of the magnets relevant for the injection optimization process are illustrated in Fig.~\ref{fig:injection_scheme}.

Since only the storage ring parameters are considered, the objective function is also chosen accordingly to decouple the effect of the pre-accelerators. The injection efficiency is defined as the net injected current into the storage ring (SR) normalized by the current measured at the booster synchrotron (S) extraction point before the beam enters the injection line to KARA
\begin{equation}
\label{eq:objective_real}
    f_\text{I-eff, exp.} = \frac{\Delta I_\text{SR}}{I_\text{S}} \frac{h_\text{SR}}{h_\text{S}},
\end{equation}
where ${h_\text{SR}}$ and ${h_\text{S}}$ are the harmonic numbers of the storage ring and the booster synchrotron, respectively. The change of the storage ring current $\Delta I_\text{SR}$ and the extracted bunch current from the synchrotron $I_\text{S}$ are read out synchronously to calculate the objective function.
Whereas for the simulation study, $n_i$ electrons extracted from the booster synchrotron and $n_s$ already stored electrons in the storage ring are tracked. The objective function is then calculated from the electrons that are successfully injected into the storage ring $n_{i,\text{success}}$ and the stored electrons that are lost during the injection process $n_{s,\text{lost}}$
\begin{equation}
\label{eq:objective_sim}
    f_\text{I-eff, sim.} = \frac{n_{i,\text{success}}-n_{s,\text{lost}}}{n_i}.
\end{equation}

During a realistic beam injection process at a storage ring, the stored current is usually much higher than the injected bunch current. Thus, a detuned parameter setting causing a beam loss can result in a very large negative objective function value and distort the landscape of the GP model.
Additionally, there could be outliers due to faulty readback values or occasional beam losses. 
In order to mitigate these effects, the lower limit of the injection efficiency is constrained to be $-1$, so that the GP model is more robust to the fluctuation of readback values. 
This should not affect the parameter region near the optima, where the stored beam is minimally disturbed and the injection efficiency is always positive.

The BO package implemented in this study uses the Python interface \textit{pyepics}~\cite{pyepics} for the communication with the control system of the accelerators, which is based on the Experimental Physics and Industrial Control Systems (EPICS)~\cite{Dalesio1991}. The software package \textit{GPy}~\cite{gpy2014} is used for building the GP model and \textit{scipy}~\cite{2020SciPy-NMeth} functionalities are used for maximizing the acquisition functions.

\subsection{\label{sec:simulation}Optimization Results in Simulation}

Before deploying the algorithms at the accelerators, the BO is first tested on a simulation. 
The injection model is built in Accelerator Toolbox for MATLAB (AT)~\cite{Terebilo2001}. We used a simplified lattice without magnet errors based on an existing MATLAB model for KARA~\cite{Portmann2005,Marsching2011}.

\begin{figure}[tb]
\centering
\includegraphics[width=\columnwidth]{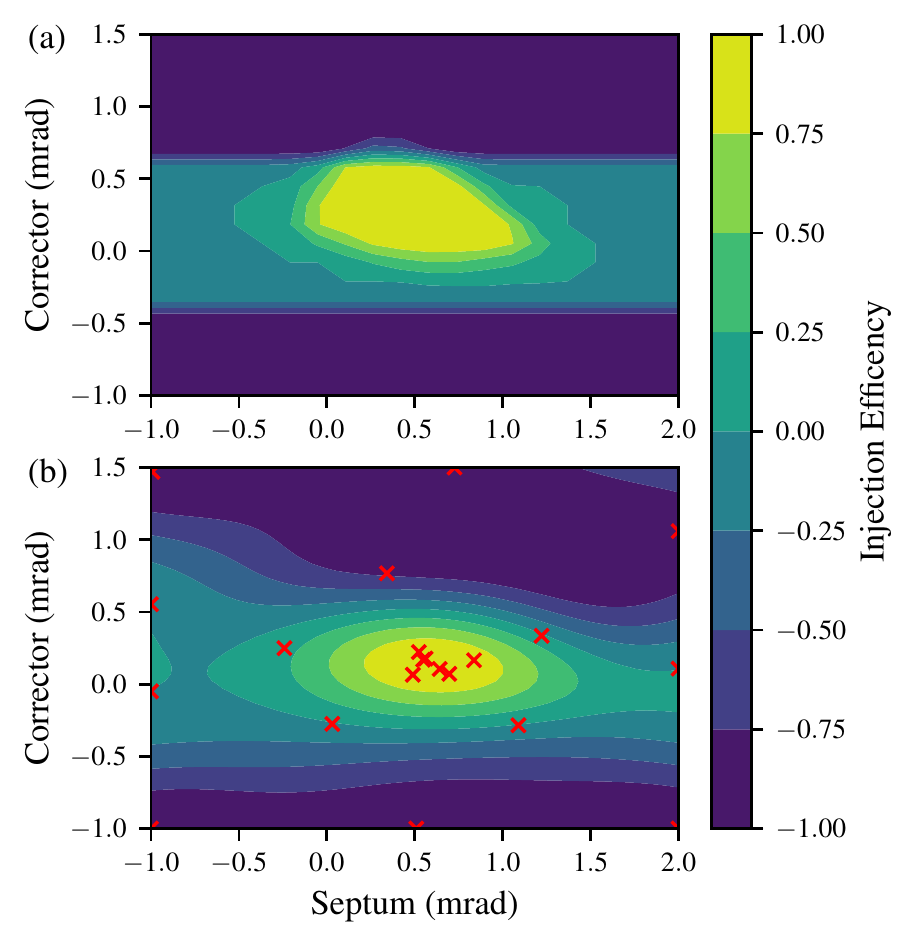}
\caption{\label{fig:bo_grid} Comparison of (a) a 2-dimensional grid scan (grid size 400) of the injection efficiency with (b) the GP posterior mean predicted based on 20 BO evaluation steps, shown in red crosses. The tuning parameters are the strengths of the horizontal corrector magnet and the septum magnet, given in deflection angles.
The BO can approximate the simple structure of parameter space in a small number of evaluation steps.}
\end{figure}

In the simulation, $n_i$ injected and $n_s$ stored electrons are generated according to the particle distribution and tracked for the first 100 turns after injection without collective effects. This proved to be sufficient, as over 95\% of the particle loss happens during the first 10 turns. The injection efficiency is calculated according to Eq.~\ref{eq:objective_sim}.

The BO algorithm is first tested using two tuning parameters: the septum and the corrector magnet strength. For comparison, a grid scan, shown in Fig.~\ref{fig:bo_grid}a, is conducted with 400 observations in total. Fig.~\ref{fig:bo_grid}b shows the GP posterior mean function after 20 evaluations. It can be seen that the BO already converges to the maximum and is able to model the peak structure.

\begin{figure}[tb]
\centering
\includegraphics[width=\columnwidth]{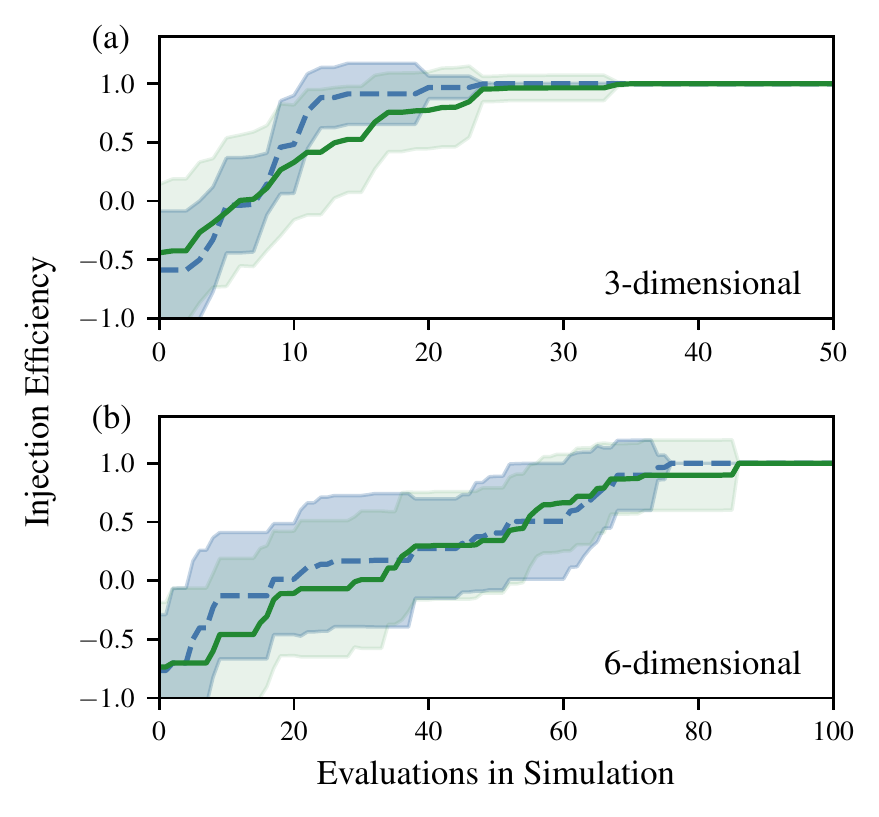}
\caption{\label{fig:sim_result} Optimization results in an AT simulation model using UCB (green, solid) and EI (blue, dashed) acquisition functions for the (a) 3- and (b) 6-dimensional problems. The lines depict the best evaluated injection efficiency in each run, averaged over 10 independent runs. The shaded areas are the one $\sigma$ spread. Within 50 and 100 steps for 3- and 6-dimensional problems respectively, all the optimization runs converged to the optimum setting. The required steps for convergence increases with the input dimensions and both acquisition functions have a similar performance.}
\end{figure}

As a next step, the RF frequency is included as a third tuning parameter. For the optimization, the BO randomly samples 5 points to initialize the GP model. The results on the 3-dimensional problem using two different acquisition functions are shown in Fig.~\ref{fig:sim_result}a. It can be seen that both EI and UCB are able to find the optimum in a small number of evaluation steps.
Then, we include the strength of the three kicker magnets as tuning parameters and the results are shown in Fig.~\ref{fig:sim_result}b. Although the number of required evaluations is larger, BO can still efficiently solve the 6-dimensional problem.

\begin{figure}[tb]
\centering
\includegraphics[width=\columnwidth]{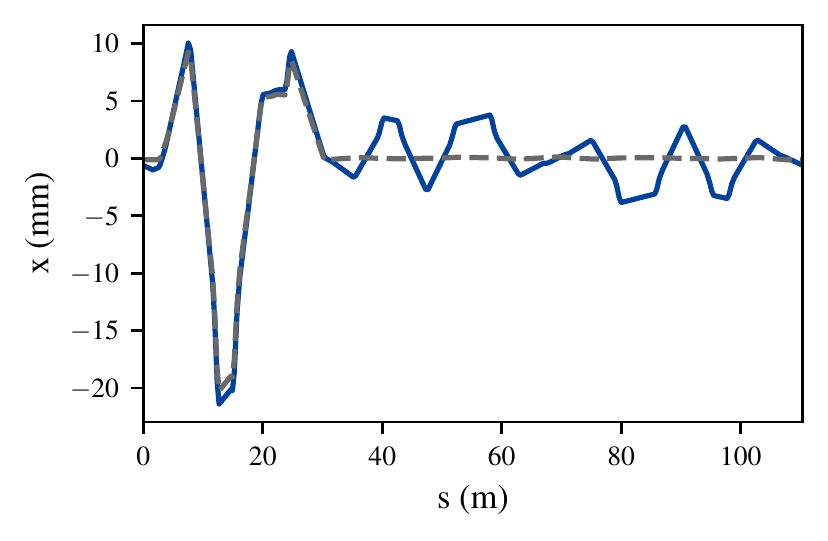}
\caption{\label{fig:orbit} Example of a non-localized orbit defined by the magnet strengths found by the BO algorithm (navy, solid), in comparison to the reference orbit (grey, dashed) with design values, where the injection bump orbit is fully localized. The orbit oscillation outside of the injection bump is not taken into consideration by BO, as it causes no beam loss.}
\end{figure}

Nevertheless, BO does not take other physical properties into account, since it only optimizes for the objective function $f_\text{I-eff, sim.}$ (cf.~Eq.\ref{eq:objective_sim}). For example, as can be seen in Fig.~\ref{fig:orbit}, the strength of the three kicker magnets is often not matched perfectly, which leads to a non-localized injection bump orbit. Although a large orbit oscillation of about \SI{3}{mm} along the storage ring is visible, the electrons can be stored and no beam loss is observed.

\subsection{\label{sec:pre}Beam Lifetime Correction}

% motivation; physical explanation;

In simulation, the collective effects are neglected and the beam losses are only due to detuned settings of the injection magnets, while in the actual storage ring the electrons are lost over time due to other effects such as scattering within the bunch and current dependent tune-shifts.
In this study, we focus on the Touschek effect, which is one of the main contributions to the reduction of beam lifetime. Additionally, particles can be lost due to the scattering with the residual gas, which depends on the vacuum quality. The two effects can be combined in terms of the total lifetime $\tau$
\begin{equation}
\label{eq:touschek}
    \frac{1}{\tau} = \frac{1}{\tau_\text{T}} + \frac{1}{\tau_\text{other}} =  aI+b,
\end{equation}
where the Touschek lifetime $\tau_\text{T}$ is dominant and inversely proportional to the current $I$, and other contributions $\tau_\text{other}$, including beam-gas scattering, are approximately constant.

The beam lifetime decreases for higher accumulated current and leads to a decrease of the calculated injection efficiency, which is independent of the performance of the BO algorithm. 
As a result, the objective function will continue to decrease regardless of the sampling region, which is not taken into account in our definition of the objective. It is observed that BO often fails to optimize, if this effect is not properly dealt with.

This effect can be counteracted by including the beam current as a context variable, which is discussed in section~\ref{sec:cbo_result}. Alternatively, we explicitly correct this effect for normal BO by measuring the lifetime related beam loss.
For the latter, we measured the beam lifetime at the injection optics and fitted the Eq.~\ref{eq:touschek} to it. The injection efficiencies are then corrected accordingly in the next experiments.

\subsection{\label{sec:result}Experimental Optimization Results}

In the following section, we describe the results of the BO algorithms in a real-world accelerator environment. We investigate its performance using the EI and UCB acquisition functions introduced in Section \ref{sec:acq}. 

\begin{figure}[tbh]
\centering
\includegraphics[width=\columnwidth]{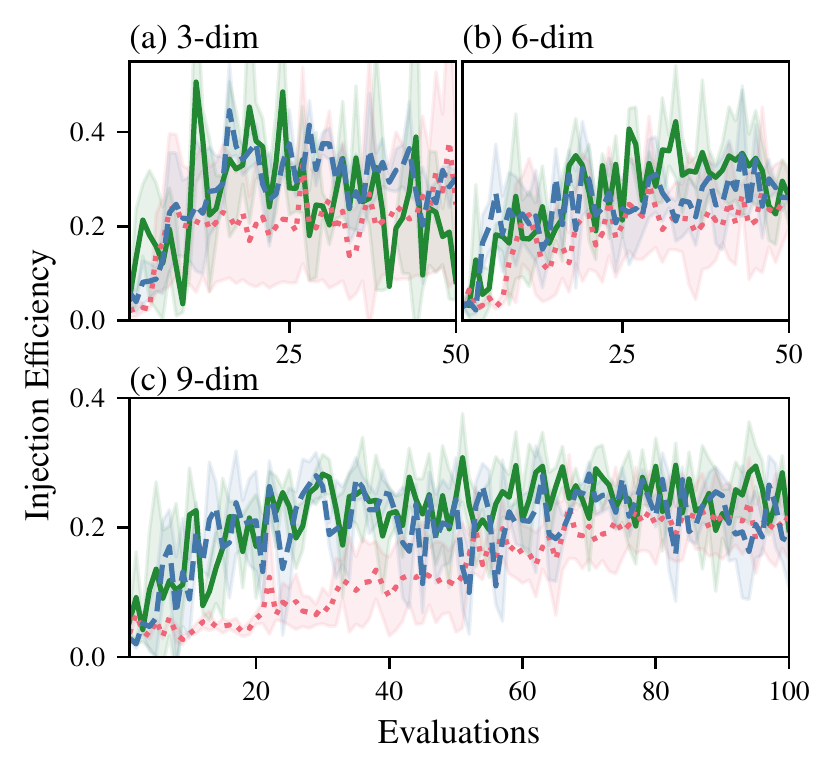}
\caption{\label{fig:results} 
Experimental optimization result of the injection efficiency using BO with two different acquisition functions and Nelder-Mead algorithm (red, dotted). UCB (green, solid) and EI (blue, dashed) have similar results and are both able to optimize the injection efficiency within a reasonable number of evaluations. For 9-dimensional problem, NM is clearly slower than BO.
}
\end{figure}

We use the RBF kernel (Eq.~\ref{eq:rbf_kernel}) with the hyperparameters determined from the results in Section~\ref{sec:gp}. For each test run, the machine was manually detuned to a fixed initial setting, where the beam injection was still possible with a very low injection efficiency. For the UCB acquisition function shown in Eq.~\ref{eq:UCB}, we chose the setting of $\kappa$ according to~\cite{Srinivas2010}, which dynamically changes the exploration-exploitation behavior
\begin{equation}
    \kappa=\sqrt{2\nu\log{(t^{d/2+2}\pi^{2}/(3\delta)}},
\end{equation}
with $\nu=1$ and $\delta=0.1$ being free parameters. The parameter $\kappa$ becomes larger with the number of evaluations $t$ and the algorithm focuses more on the exploration. 

The optimization was performed for a fixed number of steps: 50 steps for 3- and 6-dimensional problems and 100 steps for 9-dimensional problem. This corresponds to about 10-20 minutes for each optimization run, where most of the time was spent on setting the parameter values and the evaluating the injection efficiency. The computation time of the GP model and the acquisition function is negligible. 
Additionally, it should be noted that we intentionally decreased the total booster synchrotron current $I_\text{S}$ in the following experiments to reduce the radiation dose. This is expected not to influence the result, as the objective $f_\text{I-eff, exp.}$ defined in Eq.~\ref{eq:objective_real} is normalized with respect to $I_\text{S}$.

Figure~{\ref{fig:results}} shows the averaged performance over 3 optimization runs for each setting.
For all the optimization runs, the injection efficiency clearly increased from the detuned initial setting within the allowed evaluation steps.
The final achieved objective function value is, however, lower than the result obtained in simulation studies, which is due to various reasons. First, the objective function $f_\text{I-eff, exp}$ is normalized with respect to the current measured in booster synchrotron $I_\text{S}$ right before extraction. The beam loss in the injection line is not accounted for, which needs to be mitigated by including additional magnets in BO. However this is beyond the scope of this paper.
Secondly, the current readout in the storage ring and the synchrotron booster are calibrated independently, affecting their ratio.
Nevertheless, the performance of the BO can be compared and benchmarked qualitatively.
It can be seen that in the 3-dimensional case (Fig.~{\ref{fig:results}}a), the UCB evaluations are more noisy than the other configurations, which is mainly due to the setting of the trade-off parameter $\kappa$. However, for 6- and 9-dimensional problems, the parameter space is larger and more exploration becomes indeed necessary. Therefore, the performance of both UCB and EI is quite similar in those cases (Fig.~{\ref{fig:results} b,c}).
Due to its focus on exploration, occasional unexpected beam loss is observed during the UCB runs. This can be circumvented either by introducing a safety constraint, or setting $\kappa$ to a lower value to focus more on exploitation.

For comparison, we perform the same optimization procedure using the Nelder-Mead algorithm~\cite{Nelder1965}. Nelder-Mead is a widely used numerical optimization method and is considered as a standard benchmark in the accelerator community~\cite{Huang2018}. The algorithm keeps track of $d+1$ evaluation points and uses these to build a simplex. The Nelder-Mead method searches for the optimum via geometric modifications of the stored simplex and is proven to converge relatively fast.
The results of the Nelder-Mead algorithm are also plotted in Fig.~\ref{fig:results}. For the 3- and 6-dimensional cases, Nelder-Mead achieves a similar performance to BO, whereas for the 9 dimensional problem the optimization speed of Nelder-Mead is clearly slower. The fact that Nelder-Mead scales not as well as BO is expected, as the local search becomes less applicable for higher dimensional spaces. The primary problem for BO to scale to higher dimensions is the computation time of the GP model and maximizing the acquisition function. However, these are still noncritical for $d<20$~\cite{Frazier2018}.
It is worth mentioning that we used randomly selected parameter settings to initialize the GP model, which often leads to a poor performance in the first few steps, as can be seen in Fig.~\ref{fig:results} 3-dimensional case. 
Alternatively, one can mitigate that by using the initial steps of other algorithms, such as Nelder-Mead, or some historical data to initialize the GP model. 
This could avoid the unstable phase of BO in the beginning and subsequently reach quicker convergence.

\begin{figure}[tb]
\centering
\includegraphics[width=\columnwidth]{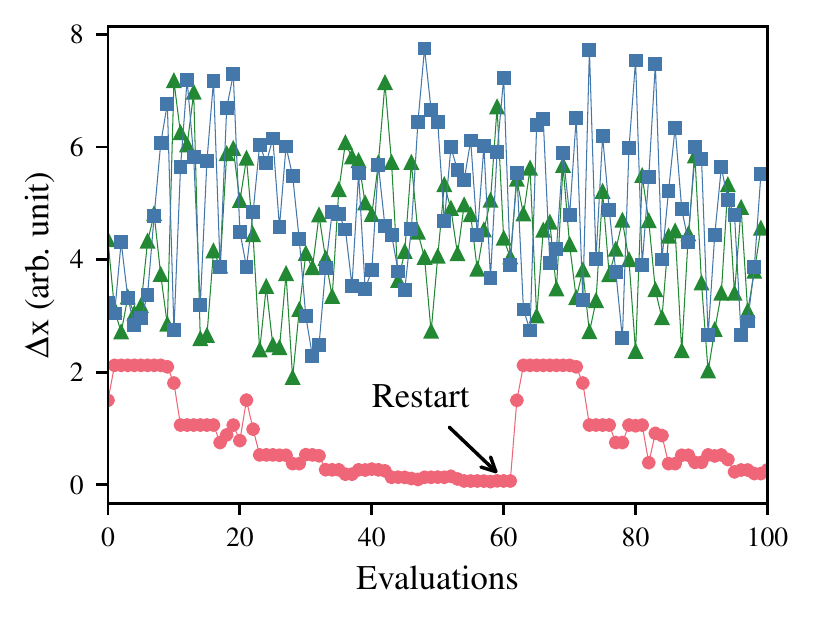}
\caption{\label{fig:results_distance} Convergence behavior of BO with UCB (green, triangle) and EI (blue, square) acquisition functions and Nelder-Mead (red, circle) on the 9-dimensional problem showing the distance $\Delta x$ between two consecutive sampled points in the scaled parameter space. Due to the noisy evaluation of the objective function, Nelder-Mead optimizes mostly locally and often needs to be restarted, whereas BO optimizes globally and doesn't require a restart.}
\end{figure}

The distances $\Delta x$ between two consecutive sampled points of BO and Nelder-Mead for the 9-dimensional problem are shown in Fig.~\ref{fig:results_distance}. It is visible that Nelder-Mead searches only locally. Since Nelder-Mead cannot handle the noisy signal well, it easily gets stuck and continues to contract, which eventually converges to a non-optimal setting. 
To prevent Nelder-Mead from breaking down and force exploration of the parameter space, it is automatically restarted when the step becomes too small. 
On the contrary, BO is more robust against the observation noise by explicitly modeling it as a hyperparameter. It can be seen that BO generally samples at a larger distance and does not become trapped in local optima.

In the experiments we found that the timing of the the kicker magnets are consistent with the previous set values. As visible in Fig.~{\ref{fig:results}c}, adding the timing parameters only slows down the optimization and does not improve the objective function further. Thus, we run the BO always with 6 parameters in the following sections.

As mentioned in the Section~\ref{sec:simulation}, the settings found by BO approach sometimes result in a not closed injection bump orbit. We also observed this effect during the experiments on KARA, despite the fact that the orbit oscillation is partly mitigated by the radiation damping. Since the trajectory of the electron beam is not centered through the quadrupole and sextupole magnets, the resulting orbit is often different from the one usually obtained via manual tuning and sometimes a slight tune shift is visible.
Although this effect does not affect the beam injection process for KARA, it might be critical for other accelerators with more stringent constraints on orbit and betatron tunes.

\begin{figure}[tb]
\includegraphics[width=\columnwidth]{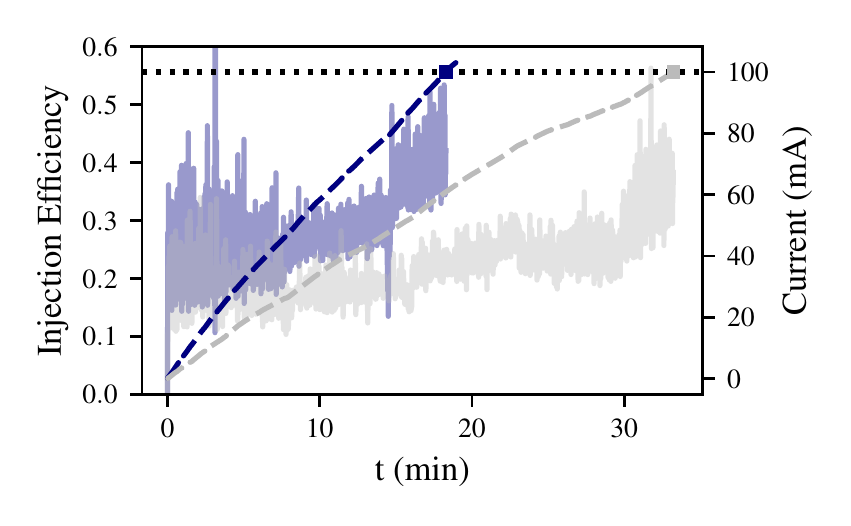}
\caption{\label{fig:injection_time} Beam injection up to \SI{100}{mA} using BO result (navy) compared to the manual tuning result (grey) obtained from an operation day. The dashed lines are the accumulated current of BO and manual tuning result. BO was roughly twice as fast as the manual tuning, greatly reducing the required beam injection time.}
\end{figure}

Lastly, the optimal settings found by BO are compared to the result of one day with dedicated manual tuning, taken a week prior to the BO experiment with a comparable accelerator condition. In this experiment, the full booster current is used in order to achieve higher injection rate. As shown in Fig.~\ref{fig:injection_time}, BO took \SI{18}{min} and manual tuning took \SI{33}{min} for the injection of \SI{100}{mA} current.
BO clearly outperforms the manual tuning and significantly reduces the required time for beam injection.

\subsection{\label{sec:new_magnet}Application to the Commissioning of Injection Magnets}

% Here, we demonstrate a use case of the Bayesian optimization for the a commissioning task.
During a scheduled shutdown period, the power supplies of the septum and kicker magnets were exchanged. Thus, it was necessary to find new parameter settings during commissioning.

\begin{figure}[tb]
\includegraphics[width=\columnwidth]{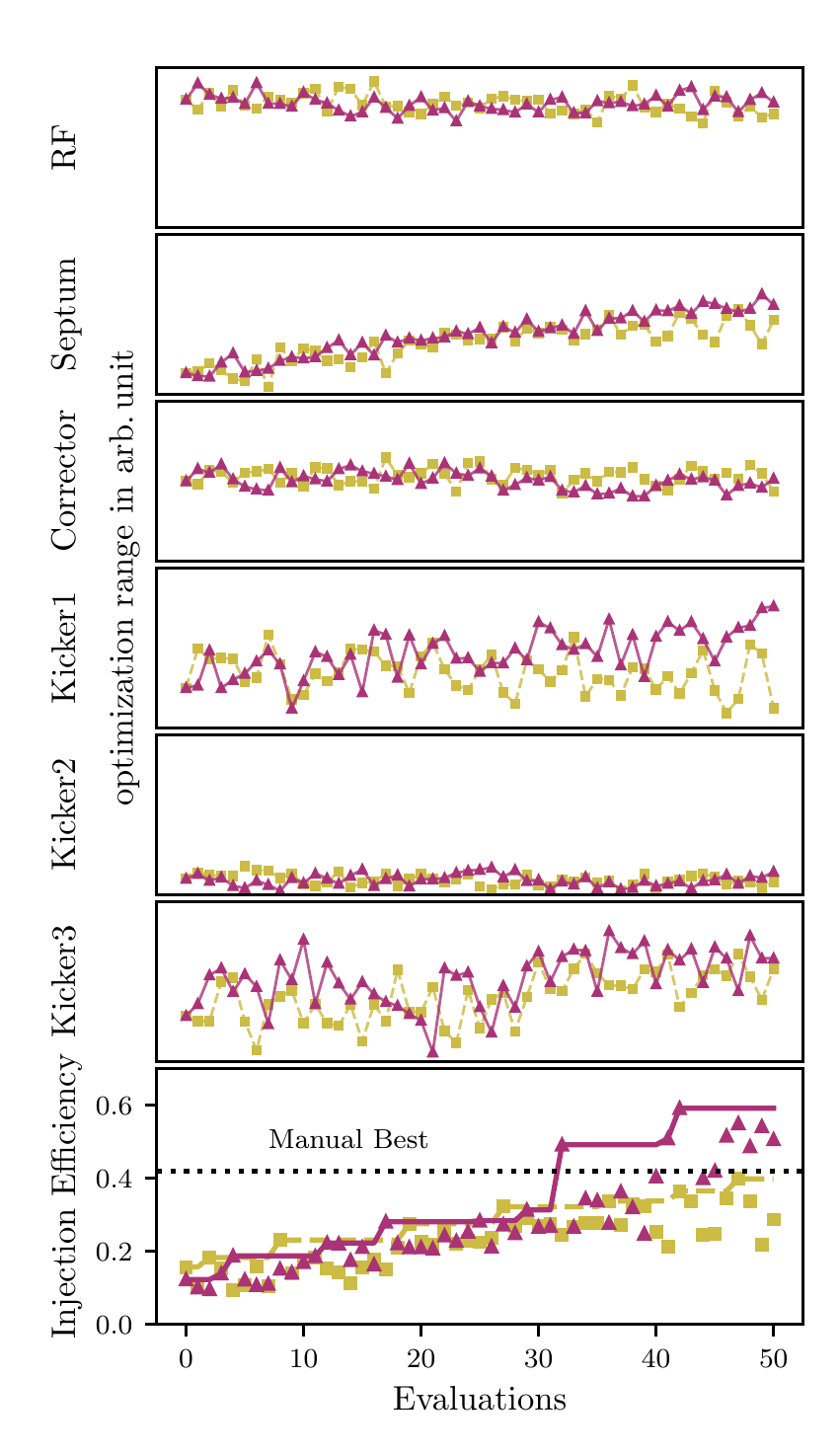}
\caption{\label{fig:commissioning_result} Two runs with the highest (purple, solid) and lowest (yellow, dashed) optimized injection efficiency achieved are plotted out of all five BO runs with UCB acquisition using the new power supplies. The evolution of the input parameters and the resulting objective function are shown over the optimization steps, where the vertical limit is the total allowed optimization range respectively. The bottom plot shows the measured injection functions (markers) with the cumulative best results (lines). The dotted line depicts the 90\% quantile of the injection efficiency obtained by manual operation in a month prior to the shutdown period. Even the BO run with worst performance achieved a comparable result as the manual baseline.}
\end{figure}

Two steps are needed to apply BO and assist the commissioning of the new power supplies: 1.~The interface with the machine is updated, so that BO can control the parameters of the new magnets. 2.~Several parameter scans are conducted in order to estimate the length scales and optimization ranges for the new parameters.
Once BO was reconfigured for the new machine status, it quickly found new working points for the beam injection.

The BO started from a fixed non-optimal setting and was run multiple times using 6 input parameters. Figure~\ref{fig:commissioning_result} compares two of these runs with the highest and lowest achieved injection efficiency among all five runs. The evolution of the parameters during the optimization are shown within the allowed tuning range of each parameter respectively.
The RF frequency is varied very little and the set values are comparable for two runs, which is as expected. Since the RF frequency directly depends on the ring circumference, it should only change due to, for instance, the change of ambient temperature over the year.
The optimized septum strength is clearly higher than the starting point. This is consistent with the expectation that the septum magnet strongly influences the injection process, as it directly acts on the injected electron bunch. 
On the other hand, the strength of kicker 2 barely changed. Since the kicker magnet 2 is located closely after the septum magnet, it possibly implies that the BO prefers to use only the septum magnet for beam deflection.
The two runs shown here have found different values for kicker 1 and 3, which lead to different bump orbit shape and eventually different injection efficiency. 

We also show the manual performance benchmark in Fig.~\ref{fig:commissioning_result}, where the data is collected in a month of synchrotron light operation prior to the shutdown period with storage ring current up to \SI{100}{mA}. This value $f_\text{manual} = 0.43$ is calculated as the 90\% quantile of the injection efficiencies obtained by manual tuning, so that the outliers due to the fluctuation of readback values are excluded. 
All the BO runs reached or exceeded the optimal settings found by manual tuning. This result shows that BO is robust against the change of the accelerator's condition and can indeed reduce the commissioning time after even drastic changes to the accelerator environment such as replacing elements to new model components.

\subsection{\label{sec:cbo_result}Contextual Optimization}

Lastly, we present the first results of including contextual parameters and extending the BO algorithm to contextual Bayesian optimization (CBO)~\cite{Krause2011}.

\begin{figure}[tb]
\includegraphics[width=\columnwidth]{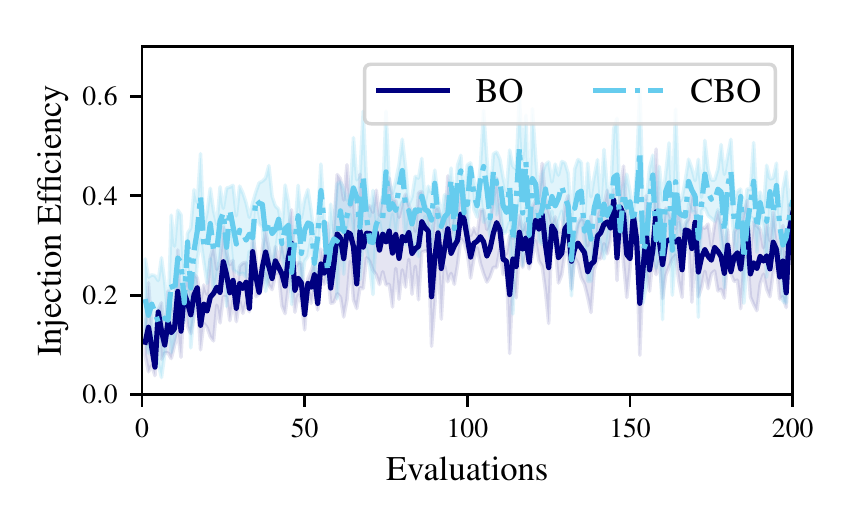}
\caption{\label{fig:benchmark_cbo} Optimization results of BO (navy, solid) and contextual BO (cyan, dashdot) on the 6-dimensional problem. Each line is the averaged injection efficiencies from three runs. By using the stored current as a context variable, CBO optimizes faster and obtains a consistently higher injection efficiency than normal BO after about 10 steps.}
\end{figure}

In the above experiments we treat the beam injection problem as stationary, which is only true as an approximation. In reality, the optimal injection condition is affected by various context variables, which are ignored by the normal BO. One example are collective effects, which become more relevant with the accumulated current. For instance, a parameter setting with a larger closed orbit is non-critical for low bunch current and could be considered good by the normal BO. However, the same setting might cause more instability and become non-optimal for a higher current.
Therefore, we include the storage ring current $I$ as a context variable by adding the current as an additional dimension to the RBF kernel and the GP model. The acquisition function is then maximized over the parameter space with the context dimension fixed.  
In this case, we set the length-scale of the current to be $l_{I}=\SI{5}{mA}$ by empirical testing. We do not correct the loss rate explicitly as in normal BO, and let the algorithm itself recognize this effect.
As before, we initialize CBO from a manually detuned setting and use UCB acquisition. The number of evaluation steps is set to 200, allowing the optimizer to achieve a higher current value. The normal BO is performed with the same configurations. Figure~\ref{fig:benchmark_cbo} shows the results, where each line is the mean objective averaged from three runs. The CBO results are corrected afterwards for the comparison with BO results.
We can see that CBO can properly handle the current-dependent effect and optimizes efficiently. It also shows a consistently better performance than BO. This is partly due to the fact that the context variable provides an extra dimension for exploration. To some extent, CBO is able to track the optimal setting along the current dimension.

\section{Summary and Outlook}

In this article, we presented the implementation of Bayesian optimization applied to the injection task at a storage ring. The algorithms were tested in simulation and also in experiments, for which the accelerator test facility and storage ring KARA at KIT was used. 
The hyperparameters were determined from some simple scan measurements, since no dedicated archive data with diverse parameter settings was available.
We showed that BO can optimize the injection efficiency in a small number of evaluations, which makes it suitable for online optimization. We demonstrated experimentally that the BO method is robust against noisy observations, outperforming the commonly used Nelder-Mead method. 
We also present a real use case of BO for machine commissioning and show that BO can be quickly reconfigured to find new parameter settings after a change in the machine status.
Lastly, we use the accumulated current as a context variable for the optimization, which shows a better performance than explicitly correcting for the Touschek effect.
The final goal of this work is to fully automate the injection tuning process, where the BO algorithm implemented in this paper serves as a basic framework. It is now used by operators routinely to assist the injection optimization task for the weekly machine start-up.
Nevertheless, the current version of BO could explore excessively and lose the stored beam due to a poor setting of the kicker magnets. To mitigate that, safety constraints can be included, so that the exploration is restricted to safe parameter regions. With safety constraints implemented, BO is expected to be also applicable for more use cases, or even at other accelerators with a more stringent machine protection requirements.

The required computation time for the presented BO with 9 tuning parameters is still negligible compared to the setting of new accelerator parameters. Thus we expect that this framework can be effortlessly extended to a few tens of tuning parameters by, for example, adding all magnets in the pre-accelerators. Other context parameters, such as temperature and vacuum quality, can also be incorporated.

Optimizing only for the highest injection efficiency sometimes results in an unmatched setting of the kicker magnets where the injection bump is not localized, which  changes the closed orbit and can lead to a tune shift or a lower beam lifetime.
This can be mitigated by considering the beam orbit and betatron tune as additional objective functions.
Generalizing the single objective BO to multi-objective optimization makes it possible to simultaneously trade-off between multiple objectives, like the orbit deviation and the injection rate, and find optimal parameter settings for user-specific cases.
Furthermore, the BO approach may find its application in other accelerator control tasks. It is best-suited for quasi-static tuning tasks, where the optimization condition is not drastically changed within the time scale of the optimization and repeated evaluation of same settings will result in comparable objective function values, e.g.\ tuning a linear accelerator or a free-electron laser (FEL).
% Add another discussion on BO improvement?
Using BO in tandem with other tuning methods will further improve its performance. For example, the Nelder-Mead can be used to provide initial samples of the GP model, and the extremum seeking algorithm can be used as a feedback system to track the optimum obtained by BO.

Whereas for solving dynamic problems for which the accelerator condition continues to change during the optimization, we expect reinforcement learning (RL) to be a promising alternative method \cite{StJohn2021, Eichler2021, Boltz2019, Wang2021}.
We believe that combining the strength of BO and RL could automate the accelerator operation to a great extent and achieve performance beyond manual-tuning.

\section*{Acknowledgements}
% maybe KSETA here?
We thank E.~Bründermann for reviewing and providing invaluable suggestions for improving this paper.
C.~Xu acknowledges the support by the DFG-funded Doctoral School "Karlsruhe School of Elementary and Astroparticle Physics: Science and Technology".

\bibliography{bayesopt}% Produces the bibliography via BibTeX.

\begin{thebibliography}{37}%
\makeatletter
\providecommand \@ifxundefined [1]{%
 \@ifx{#1\undefined}
}%
\providecommand \@ifnum [1]{%
 \ifnum #1\expandafter \@firstoftwo
 \else \expandafter \@secondoftwo
 \fi
}%
\providecommand \@ifx [1]{%
 \ifx #1\expandafter \@firstoftwo
 \else \expandafter \@secondoftwo
 \fi
}%
\providecommand \natexlab [1]{#1}%
\providecommand \enquote  [1]{``#1''}%
\providecommand \bibnamefont  [1]{#1}%
\providecommand \bibfnamefont [1]{#1}%
\providecommand \citenamefont [1]{#1}%
\providecommand \href@noop [0]{\@secondoftwo}%
\providecommand \href [0]{\begingroup \@sanitize@url \@href}%
\providecommand \@href[1]{\@@startlink{#1}\@@href}%
\providecommand \@@href[1]{\endgroup#1\@@endlink}%
\providecommand \@sanitize@url [0]{\catcode `\\12\catcode `\$12\catcode
  `\&12\catcode `\#12\catcode `\^12\catcode `\_12\catcode `\%12\relax}%
\providecommand \@@startlink[1]{}%
\providecommand \@@endlink[0]{}%
\providecommand \url  [0]{\begingroup\@sanitize@url \@url }%
\providecommand \@url [1]{\endgroup\@href {#1}{\urlprefix }}%
\providecommand \urlprefix  [0]{URL }%
\providecommand \Eprint [0]{\href }%
\providecommand \doibase [0]{https://doi.org/}%
\providecommand \selectlanguage [0]{\@gobble}%
\providecommand \bibinfo  [0]{\@secondoftwo}%
\providecommand \bibfield  [0]{\@secondoftwo}%
\providecommand \translation [1]{[#1]}%
\providecommand \BibitemOpen [0]{}%
\providecommand \bibitemStop [0]{}%
\providecommand \bibitemNoStop [0]{.\EOS\space}%
\providecommand \EOS [0]{\spacefactor3000\relax}%
\providecommand \BibitemShut  [1]{\csname bibitem#1\endcsname}%
\let\auto@bib@innerbib\@empty
%</preamble>
\bibitem [{\citenamefont {Bründermann}\ \emph {et~al.}(2012)\citenamefont
  {Bründermann}, \citenamefont {Hübers}, \citenamefont {Fitz},\ and\
  \citenamefont {Kimmitt}}]{Bruendermann2012}%
  \BibitemOpen
  \bibfield  {author} {\bibinfo {author} {\bibfnamefont {E.}~\bibnamefont
  {Bründermann}}, \bibinfo {author} {\bibfnamefont {H.-W.}\ \bibnamefont
  {Hübers}}, \bibinfo {author} {\bibfnamefont {M.}~\bibnamefont {Fitz}},\ and\
  \bibinfo {author} {\bibfnamefont {G.}~\bibnamefont {Kimmitt}},\ }\href@noop
  {} {\emph {\bibinfo {title} {Terahertz techniques}}},\ \bibinfo {series}
  {Springer series in optical sciences}, Vol.\ \bibinfo {volume} {151}\
  (\bibinfo  {publisher} {{Springer Verlag}},\ \bibinfo {year} {2012})\
  \bibinfo {note} {54.01.01; LK 01}\BibitemShut {NoStop}%
\bibitem [{\citenamefont {Bernardini}\ \emph {et~al.}(1963)\citenamefont
  {Bernardini}, \citenamefont {Corazza}, \citenamefont {Di~Giugno},
  \citenamefont {Ghigo}, \citenamefont {Haissinski}, \citenamefont {Marin},
  \citenamefont {Querzoli},\ and\ \citenamefont {Touschek}}]{Bernardini1963}%
  \BibitemOpen
  \bibfield  {author} {\bibinfo {author} {\bibfnamefont {C.}~\bibnamefont
  {Bernardini}}, \bibinfo {author} {\bibfnamefont {G.~F.}\ \bibnamefont
  {Corazza}}, \bibinfo {author} {\bibfnamefont {G.}~\bibnamefont {Di~Giugno}},
  \bibinfo {author} {\bibfnamefont {G.}~\bibnamefont {Ghigo}}, \bibinfo
  {author} {\bibfnamefont {J.}~\bibnamefont {Haissinski}}, \bibinfo {author}
  {\bibfnamefont {P.}~\bibnamefont {Marin}}, \bibinfo {author} {\bibfnamefont
  {R.}~\bibnamefont {Querzoli}},\ and\ \bibinfo {author} {\bibfnamefont
  {B.}~\bibnamefont {Touschek}},\ }\bibfield  {title} {\bibinfo {title}
  {Lifetime and beam size in a storage ring},\ }\href
  {https://doi.org/10.1103/PhysRevLett.10.407} {\bibfield  {journal} {\bibinfo
  {journal} {Phys. Rev. Lett.}\ }\textbf {\bibinfo {volume} {10}},\ \bibinfo
  {pages} {407} (\bibinfo {year} {1963})}\BibitemShut {NoStop}%
\bibitem [{\citenamefont {Edelen}\ \emph {et~al.}(2018)\citenamefont {Edelen},
  \citenamefont {Mayes}, \citenamefont {Bowring}, \citenamefont {Ratner},
  \citenamefont {Adelmann}, \citenamefont {Ischebeck}, \citenamefont
  {Snuverink}, \citenamefont {Agapov}, \citenamefont {Kammering}, \citenamefont
  {Edelen}, \citenamefont {Bazarov}, \citenamefont {Valentino},\ and\
  \citenamefont {Wenninger}}]{Edelen2018a}%
  \BibitemOpen
  \bibfield  {author} {\bibinfo {author} {\bibfnamefont {A.}~\bibnamefont
  {Edelen}}, \bibinfo {author} {\bibfnamefont {C.}~\bibnamefont {Mayes}},
  \bibinfo {author} {\bibfnamefont {D.}~\bibnamefont {Bowring}}, \bibinfo
  {author} {\bibfnamefont {D.}~\bibnamefont {Ratner}}, \bibinfo {author}
  {\bibfnamefont {A.}~\bibnamefont {Adelmann}}, \bibinfo {author}
  {\bibfnamefont {R.}~\bibnamefont {Ischebeck}}, \bibinfo {author}
  {\bibfnamefont {J.}~\bibnamefont {Snuverink}}, \bibinfo {author}
  {\bibfnamefont {I.}~\bibnamefont {Agapov}}, \bibinfo {author} {\bibfnamefont
  {R.}~\bibnamefont {Kammering}}, \bibinfo {author} {\bibfnamefont
  {J.}~\bibnamefont {Edelen}}, \bibinfo {author} {\bibfnamefont
  {I.}~\bibnamefont {Bazarov}}, \bibinfo {author} {\bibfnamefont
  {G.}~\bibnamefont {Valentino}},\ and\ \bibinfo {author} {\bibfnamefont
  {J.}~\bibnamefont {Wenninger}},\ }\href {http://arxiv.org/abs/1811.03172}
  {\bibinfo {title} {{Opportunities in Machine Learning for Particle
  Accelerators}}} (\bibinfo {year} {2018}),\ \Eprint
  {https://arxiv.org/abs/1811.03172} {arXiv:1811.03172} \BibitemShut {NoStop}%
\bibitem [{\citenamefont {{Vikhar}}(2016)}]{Vikhar2016}%
  \BibitemOpen
  \bibfield  {author} {\bibinfo {author} {\bibfnamefont {P.~A.}\ \bibnamefont
  {{Vikhar}}},\ }\bibfield  {title} {\bibinfo {title} {Evolutionary algorithms:
  A critical review and its future prospects},\ }in\ \href@noop {} {\emph
  {\bibinfo {booktitle} {2016 International Conference on Global Trends in
  Signal Processing, Information Computing and Communication (ICGTSPICC)}}}\
  (\bibinfo {year} {2016})\ pp.\ \bibinfo {pages} {261--265}\BibitemShut
  {NoStop}%
\bibitem [{\citenamefont {{Kennedy}}\ and\ \citenamefont
  {{Eberhart}}(1995)}]{Kennedy1995}%
  \BibitemOpen
  \bibfield  {author} {\bibinfo {author} {\bibfnamefont {J.}~\bibnamefont
  {{Kennedy}}}\ and\ \bibinfo {author} {\bibfnamefont {R.}~\bibnamefont
  {{Eberhart}}},\ }\bibfield  {title} {\bibinfo {title} {Particle swarm
  optimization},\ }in\ \href@noop {} {\emph {\bibinfo {booktitle} {Proceedings
  of ICNN'95 - International Conference on Neural Networks}}},\ Vol.~\bibinfo
  {volume} {4}\ (\bibinfo {year} {1995})\ pp.\ \bibinfo {pages} {1942--1948
  vol.4}\BibitemShut {NoStop}%
\bibitem [{\citenamefont {Huang}\ and\ \citenamefont
  {Safranek}(2015)}]{Huang2015}%
  \BibitemOpen
  \bibfield  {author} {\bibinfo {author} {\bibfnamefont {X.}~\bibnamefont
  {Huang}}\ and\ \bibinfo {author} {\bibfnamefont {J.}~\bibnamefont
  {Safranek}},\ }\bibfield  {title} {\bibinfo {title} {Online optimization of
  storage ring nonlinear beam dynamics},\ }\href
  {https://doi.org/10.1103/PhysRevSTAB.18.084001} {\bibfield  {journal}
  {\bibinfo  {journal} {Phys. Rev. ST Accel. Beams}\ }\textbf {\bibinfo
  {volume} {18}},\ \bibinfo {pages} {084001} (\bibinfo {year}
  {2015})}\BibitemShut {NoStop}%
\bibitem [{\citenamefont {Huang}(2018)}]{Huang2018}%
  \BibitemOpen
  \bibfield  {author} {\bibinfo {author} {\bibfnamefont {X.}~\bibnamefont
  {Huang}},\ }\bibfield  {title} {\bibinfo {title} {{Robust simplex algorithm
  for online optimization}},\ }\bibfield  {journal} {\bibinfo  {journal} {Phys.
  Rev. Accel. Beams}\ }\textbf {\bibinfo {volume} {21}},\ \href
  {https://doi.org/10.1103/PhysRevAccelBeams.21.104601}
  {10.1103/PhysRevAccelBeams.21.104601} (\bibinfo {year} {2018})\BibitemShut
  {NoStop}%
\bibitem [{\citenamefont {Scheinker}\ \emph {et~al.}(2018)\citenamefont
  {Scheinker}, \citenamefont {Huang},\ and\ \citenamefont
  {Wu}}]{Scheinker2018minimization}%
  \BibitemOpen
  \bibfield  {author} {\bibinfo {author} {\bibfnamefont {A.}~\bibnamefont
  {Scheinker}}, \bibinfo {author} {\bibfnamefont {X.}~\bibnamefont {Huang}},\
  and\ \bibinfo {author} {\bibfnamefont {J.}~\bibnamefont {Wu}},\ }\bibfield
  {title} {\bibinfo {title} {Minimization of betatron oscillations of electron
  beam injected into a time-varying lattice via extremum seeking},\ }\href
  {https://doi.org/10.1109/TCST.2017.2664728} {\bibfield  {journal} {\bibinfo
  {journal} {IEEE Transactions on Control Systems Technology}\ }\textbf
  {\bibinfo {volume} {26}},\ \bibinfo {pages} {336} (\bibinfo {year}
  {2018})}\BibitemShut {NoStop}%
\bibitem [{\citenamefont {Scheinker}\ \emph {et~al.}(2019)\citenamefont
  {Scheinker}, \citenamefont {Bohler}, \citenamefont {Tomin}, \citenamefont
  {Kammering}, \citenamefont {Zagorodnov}, \citenamefont {Schlarb},
  \citenamefont {Scholz}, \citenamefont {Beutner},\ and\ \citenamefont
  {Decking}}]{Scheinker2019model}%
  \BibitemOpen
  \bibfield  {author} {\bibinfo {author} {\bibfnamefont {A.}~\bibnamefont
  {Scheinker}}, \bibinfo {author} {\bibfnamefont {D.}~\bibnamefont {Bohler}},
  \bibinfo {author} {\bibfnamefont {S.}~\bibnamefont {Tomin}}, \bibinfo
  {author} {\bibfnamefont {R.}~\bibnamefont {Kammering}}, \bibinfo {author}
  {\bibfnamefont {I.}~\bibnamefont {Zagorodnov}}, \bibinfo {author}
  {\bibfnamefont {H.}~\bibnamefont {Schlarb}}, \bibinfo {author} {\bibfnamefont
  {M.}~\bibnamefont {Scholz}}, \bibinfo {author} {\bibfnamefont
  {B.}~\bibnamefont {Beutner}},\ and\ \bibinfo {author} {\bibfnamefont
  {W.}~\bibnamefont {Decking}},\ }\bibfield  {title} {\bibinfo {title}
  {Model-independent tuning for maximizing free electron laser pulse energy},\
  }\href {https://doi.org/10.1103/PhysRevAccelBeams.22.082802} {\bibfield
  {journal} {\bibinfo  {journal} {Phys. Rev. Accel. Beams}\ }\textbf {\bibinfo
  {volume} {22}},\ \bibinfo {pages} {082802} (\bibinfo {year}
  {2019})}\BibitemShut {NoStop}%
\bibitem [{\citenamefont {Jones}(2001)}]{Jones2001}%
  \BibitemOpen
  \bibfield  {author} {\bibinfo {author} {\bibfnamefont {D.~R.}\ \bibnamefont
  {Jones}},\ }\bibfield  {title} {\bibinfo {title} {{A Taxonomy of Global
  Optimization Methods Based on Response Surfaces}},\ }\bibfield  {journal}
  {\bibinfo  {journal} {J. Glob. Optim.}\ }\href
  {https://doi.org/10.1023/A:1012771025575} {10.1023/A:1012771025575} (\bibinfo
  {year} {2001})\BibitemShut {NoStop}%
\bibitem [{\citenamefont {Rasmussen}\ and\ \citenamefont
  {Williams}(2005)}]{Rasmussen2006}%
  \BibitemOpen
  \bibfield  {author} {\bibinfo {author} {\bibfnamefont {C.~E.}\ \bibnamefont
  {Rasmussen}}\ and\ \bibinfo {author} {\bibfnamefont {C.~K.~I.}\ \bibnamefont
  {Williams}},\ }\href {http://www.gaussianprocess.org/gpml/} {\emph {\bibinfo
  {title} {{Gaussian Processes for Machine Learning}}}}\ (\bibinfo  {publisher}
  {The MIT Press},\ \bibinfo {year} {2005})\BibitemShut {NoStop}%
\bibitem [{\citenamefont {McIntire}\ \emph {et~al.}(2016)\citenamefont
  {McIntire}, \citenamefont {Cope}, \citenamefont {Ratner},\ and\ \citenamefont
  {Ermon}}]{McIntire2016}%
  \BibitemOpen
  \bibfield  {author} {\bibinfo {author} {\bibfnamefont {M.}~\bibnamefont
  {McIntire}}, \bibinfo {author} {\bibfnamefont {T.}~\bibnamefont {Cope}},
  \bibinfo {author} {\bibfnamefont {D.}~\bibnamefont {Ratner}},\ and\ \bibinfo
  {author} {\bibfnamefont {S.}~\bibnamefont {Ermon}},\ }\bibfield  {title}
  {\bibinfo {title} {{Bayesian optimization of FEL performance at LCLS}},\ }in\
  \href@noop {} {\emph {\bibinfo {booktitle} {IPAC 2016 - Proc. 7th Int. Part.
  Accel. Conf.}}}\ (\bibinfo {year} {2016})\ pp.\ \bibinfo {pages}
  {2972--2975}\BibitemShut {NoStop}%
\bibitem [{\citenamefont {Duris}\ \emph {et~al.}(2020)\citenamefont {Duris},
  \citenamefont {Kennedy}, \citenamefont {Hanuka}, \citenamefont {Shtalenkova},
  \citenamefont {Edelen}, \citenamefont {Baxevanis}, \citenamefont {Egger},
  \citenamefont {Cope}, \citenamefont {McIntire}, \citenamefont {Ermon},\ and\
  \citenamefont {Ratner}}]{Duris2020}%
  \BibitemOpen
  \bibfield  {author} {\bibinfo {author} {\bibfnamefont {J.}~\bibnamefont
  {Duris}}, \bibinfo {author} {\bibfnamefont {D.}~\bibnamefont {Kennedy}},
  \bibinfo {author} {\bibfnamefont {A.}~\bibnamefont {Hanuka}}, \bibinfo
  {author} {\bibfnamefont {J.}~\bibnamefont {Shtalenkova}}, \bibinfo {author}
  {\bibfnamefont {A.}~\bibnamefont {Edelen}}, \bibinfo {author} {\bibfnamefont
  {P.}~\bibnamefont {Baxevanis}}, \bibinfo {author} {\bibfnamefont
  {A.}~\bibnamefont {Egger}}, \bibinfo {author} {\bibfnamefont
  {T.}~\bibnamefont {Cope}}, \bibinfo {author} {\bibfnamefont {M.}~\bibnamefont
  {McIntire}}, \bibinfo {author} {\bibfnamefont {S.}~\bibnamefont {Ermon}},\
  and\ \bibinfo {author} {\bibfnamefont {D.}~\bibnamefont {Ratner}},\
  }\bibfield  {title} {\bibinfo {title} {Bayesian optimization of a
  free-electron laser},\ }\href
  {https://doi.org/10.1103/PhysRevLett.124.124801} {\bibfield  {journal}
  {\bibinfo  {journal} {Phys. Rev. Lett.}\ }\textbf {\bibinfo {volume} {124}},\
  \bibinfo {pages} {124801} (\bibinfo {year} {2020})}\BibitemShut {NoStop}%
\bibitem [{\citenamefont {Kirschner}\ \emph {et~al.}(2019)\citenamefont
  {Kirschner}, \citenamefont {Adelmann}, \citenamefont {Hiller}, \citenamefont
  {Ischebeck}, \citenamefont {Krause}, \citenamefont {Mutn\'y},\ and\
  \citenamefont {Nonnenmacher}}]{Kirschner2019}%
  \BibitemOpen
  \bibfield  {author} {\bibinfo {author} {\bibfnamefont {J.}~\bibnamefont
  {Kirschner}}, \bibinfo {author} {\bibfnamefont {A.}~\bibnamefont {Adelmann}},
  \bibinfo {author} {\bibfnamefont {N.}~\bibnamefont {Hiller}}, \bibinfo
  {author} {\bibfnamefont {R.}~\bibnamefont {Ischebeck}}, \bibinfo {author}
  {\bibfnamefont {A.}~\bibnamefont {Krause}}, \bibinfo {author} {\bibfnamefont
  {M.}~\bibnamefont {Mutn\'y}},\ and\ \bibinfo {author} {\bibfnamefont
  {M.}~\bibnamefont {Nonnenmacher}},\ }\bibfield  {title} {\bibinfo {title}
  {{Bayesian Optimisation for Fast and Safe Parameter Tuning of SwissFEL}},\
  }in\ \href {https://doi.org/10.18429/JACoW-FEL2019-THP061} {\emph {\bibinfo
  {booktitle} {{39th International Free Electron Laser Conference}}}}\
  (\bibinfo {year} {2019})\ p.\ \bibinfo {pages} {THP061}\BibitemShut {NoStop}%
\bibitem [{\citenamefont {Hanuka}\ \emph {et~al.}(2019)\citenamefont {Hanuka},
  \citenamefont {Duris}, \citenamefont {Shtalenkova}, \citenamefont {Kennedy},
  \citenamefont {Edelen}, \citenamefont {Ratner},\ and\ \citenamefont
  {Huang}}]{Hanuka2019}%
  \BibitemOpen
  \bibfield  {author} {\bibinfo {author} {\bibfnamefont {A.}~\bibnamefont
  {Hanuka}}, \bibinfo {author} {\bibfnamefont {J.}~\bibnamefont {Duris}},
  \bibinfo {author} {\bibfnamefont {J.}~\bibnamefont {Shtalenkova}}, \bibinfo
  {author} {\bibfnamefont {D.}~\bibnamefont {Kennedy}}, \bibinfo {author}
  {\bibfnamefont {A.}~\bibnamefont {Edelen}}, \bibinfo {author} {\bibfnamefont
  {D.}~\bibnamefont {Ratner}},\ and\ \bibinfo {author} {\bibfnamefont
  {X.}~\bibnamefont {Huang}},\ }\href@noop {} {\bibinfo {title} {Online tuning
  and light source control using a physics-informed gaussian process}}
  (\bibinfo {year} {2019}),\ \Eprint {https://arxiv.org/abs/1911.01538}
  {arXiv:1911.01538 [physics.acc-ph]} \BibitemShut {NoStop}%
\bibitem [{\citenamefont {Roussel}\ \emph {et~al.}(2021)\citenamefont
  {Roussel}, \citenamefont {Hanuka},\ and\ \citenamefont
  {Edelen}}]{Roussel2020}%
  \BibitemOpen
  \bibfield  {author} {\bibinfo {author} {\bibfnamefont {R.}~\bibnamefont
  {Roussel}}, \bibinfo {author} {\bibfnamefont {A.}~\bibnamefont {Hanuka}},\
  and\ \bibinfo {author} {\bibfnamefont {A.}~\bibnamefont {Edelen}},\
  }\bibfield  {title} {\bibinfo {title} {Multiobjective bayesian optimization
  for online accelerator tuning},\ }\href
  {https://doi.org/10.1103/PhysRevAccelBeams.24.062801} {\bibfield  {journal}
  {\bibinfo  {journal} {Phys. Rev. Accel. Beams}\ }\textbf {\bibinfo {volume}
  {24}},\ \bibinfo {pages} {062801} (\bibinfo {year} {2021})}\BibitemShut
  {NoStop}%
\bibitem [{\citenamefont {Xu}(2020)}]{Xu2020}%
  \BibitemOpen
  \bibfield  {author} {\bibinfo {author} {\bibfnamefont {C.}~\bibnamefont
  {Xu}},\ }\emph {\bibinfo {title} {{Bayesian Optimization of Injection
  Efficiency at KARA using Gaussian Processes}}},\ \href@noop {} {Master's
  thesis},\ \bibinfo  {school} {Karlsruhe Institute of Technology} (\bibinfo
  {year} {2020})\BibitemShut {NoStop}%
\bibitem [{\citenamefont {Krause}\ and\ \citenamefont
  {Ong}(2011)}]{Krause2011}%
  \BibitemOpen
  \bibfield  {author} {\bibinfo {author} {\bibfnamefont {A.}~\bibnamefont
  {Krause}}\ and\ \bibinfo {author} {\bibfnamefont {C.~S.}\ \bibnamefont
  {Ong}},\ }\bibfield  {title} {\bibinfo {title} {{Contextual Gaussian process
  bandit optimization}},\ }in\ \href@noop {} {\emph {\bibinfo {booktitle} {Adv.
  Neural Inf. Process. Syst. 24 25th Annu. Conf. Neural Inf. Process. Syst.
  2011, NIPS 2011}}}\ (\bibinfo {year} {2011})\BibitemShut {NoStop}%
\bibitem [{\citenamefont {Brochu}\ \emph {et~al.}(2010)\citenamefont {Brochu},
  \citenamefont {Cora},\ and\ \citenamefont {Freitas}}]{Brochu2010}%
  \BibitemOpen
  \bibfield  {author} {\bibinfo {author} {\bibfnamefont {E.}~\bibnamefont
  {Brochu}}, \bibinfo {author} {\bibfnamefont {V.}~\bibnamefont {Cora}},\ and\
  \bibinfo {author} {\bibfnamefont {N.}~\bibnamefont {Freitas}},\ }\bibfield
  {title} {\bibinfo {title} {A tutorial on bayesian optimization of expensive
  cost functions, with application to active user modeling and hierarchical
  reinforcement learning},\ }\href@noop {} {\bibfield  {journal} {\bibinfo
  {journal} {CoRR}\ }\textbf {\bibinfo {volume} {abs/1012.2599}} (\bibinfo
  {year} {2010})}\BibitemShut {NoStop}%
\bibitem [{\citenamefont {Srinivas}\ \emph {et~al.}(2010)\citenamefont
  {Srinivas}, \citenamefont {Krause}, \citenamefont {Kakade},\ and\
  \citenamefont {Seeger}}]{Srinivas2010}%
  \BibitemOpen
  \bibfield  {author} {\bibinfo {author} {\bibfnamefont {N.}~\bibnamefont
  {Srinivas}}, \bibinfo {author} {\bibfnamefont {A.}~\bibnamefont {Krause}},
  \bibinfo {author} {\bibfnamefont {S.}~\bibnamefont {Kakade}},\ and\ \bibinfo
  {author} {\bibfnamefont {M.}~\bibnamefont {Seeger}},\ }\bibfield  {title}
  {\bibinfo {title} {{Gaussian process optimization in the bandit setting: No
  regret and experimental design}},\ }in\ \href
  {https://doi.org/10.1109/TIT.2011.2182033} {\emph {\bibinfo {booktitle} {ICML
  2010 - Proceedings, 27th Int. Conf. Mach. Learn.}}}\ (\bibinfo {year}
  {2010})\ pp.\ \bibinfo {pages} {1015--1022},\ \Eprint
  {https://arxiv.org/abs/0912.3995} {arXiv:0912.3995} \BibitemShut {NoStop}%
\bibitem [{\citenamefont {Jones}\ \emph {et~al.}(1998)\citenamefont {Jones},
  \citenamefont {Schonlau},\ and\ \citenamefont {Welch}}]{Jones1998}%
  \BibitemOpen
  \bibfield  {author} {\bibinfo {author} {\bibfnamefont {D.~R.}\ \bibnamefont
  {Jones}}, \bibinfo {author} {\bibfnamefont {M.}~\bibnamefont {Schonlau}},\
  and\ \bibinfo {author} {\bibfnamefont {W.~J.}\ \bibnamefont {Welch}},\
  }\bibfield  {title} {\bibinfo {title} {{Efficient Global Optimization of
  Expensive Black-Box Functions}},\ }\href
  {https://doi.org/10.1023/A:1008306431147} {\bibfield  {journal} {\bibinfo
  {journal} {J. Glob. Optim.}\ }\textbf {\bibinfo {volume} {13}},\ \bibinfo
  {pages} {455} (\bibinfo {year} {1998})}\BibitemShut {NoStop}%
\bibitem [{\citenamefont {Lizotte}(2008)}]{Lizotte2008}%
  \BibitemOpen
  \bibfield  {author} {\bibinfo {author} {\bibfnamefont {D.~J.}\ \bibnamefont
  {Lizotte}},\ }\emph {\bibinfo {title} {{Practical Bayesian Optimization}}},\
  \href@noop {} {Ph.D. thesis},\ \bibinfo  {school} {University of Alberta}
  (\bibinfo {year} {2008})\BibitemShut {NoStop}%
\bibitem [{\citenamefont {Einfeld}\ \emph {et~al.}(1998)\citenamefont
  {Einfeld}, \citenamefont {Hermle}, \citenamefont {Huttel}, \citenamefont
  {Rossmanith},\ and\ \citenamefont {Walther}}]{Einfeld1998}%
  \BibitemOpen
  \bibfield  {author} {\bibinfo {author} {\bibfnamefont {D.}~\bibnamefont
  {Einfeld}}, \bibinfo {author} {\bibfnamefont {S.}~\bibnamefont {Hermle}},
  \bibinfo {author} {\bibfnamefont {E.}~\bibnamefont {Huttel}}, \bibinfo
  {author} {\bibfnamefont {R.}~\bibnamefont {Rossmanith}},\ and\ \bibinfo
  {author} {\bibfnamefont {R.}~\bibnamefont {Walther}},\ }\bibfield  {title}
  {\bibinfo {title} {{The injection scheme for the ANKA storage ring}},\ }in\
  \href@noop {} {\emph {\bibinfo {booktitle} {{6th European Particle
  Accelerator Conference (EPAC 98)}}}}\ (\bibinfo {year} {1998})\ pp.\ \bibinfo
  {pages} {2135--2137}\BibitemShut {NoStop}%
\bibitem [{\citenamefont {Turner}(1994)}]{Turner:1994znj}%
  \BibitemOpen
  \bibinfo {editor} {\bibfnamefont {S.}~\bibnamefont {Turner}},\ ed.,\ \href
  {https://doi.org/10.5170/CERN-1994-001} {\emph {\bibinfo {title} {{CAS-CERN
  Accelerator School: 5th general accelerator physics course, Jyvaskyla,
  Finland, 7-18 Sep 1992: Proceedings. 2 vol.}}}},\ CERN Yellow Reports: School
  Proceedings\ (\bibinfo {year} {1994})\BibitemShut {NoStop}%
\bibitem [{\citenamefont {Newville}\ \emph {et~al.}(2019)\citenamefont
  {Newville}, \citenamefont {Lauer}, \citenamefont {dchabot}, \citenamefont
  {Caswell}, \citenamefont {Gibbs}, \citenamefont {Péteut}, \citenamefont
  {Hartman}, \citenamefont {rokvintar}, \citenamefont {Clarken}, \citenamefont
  {Martins}, \citenamefont {Allan}, \citenamefont {Birke}, \citenamefont
  {Jemian}, \citenamefont {Claesson}, \citenamefont {Adelman}, \citenamefont
  {Dwyer}, \citenamefont {Slepicka}, \citenamefont {Brandl}, \citenamefont
  {Greenberg}, \citenamefont {Vine},\ and\ \citenamefont {André}}]{pyepics}%
  \BibitemOpen
  \bibfield  {author} {\bibinfo {author} {\bibfnamefont {M.}~\bibnamefont
  {Newville}}, \bibinfo {author} {\bibfnamefont {K.}~\bibnamefont {Lauer}},
  \bibinfo {author} {\bibnamefont {dchabot}}, \bibinfo {author} {\bibfnamefont
  {T.~A.}\ \bibnamefont {Caswell}}, \bibinfo {author} {\bibfnamefont
  {M.}~\bibnamefont {Gibbs}}, \bibinfo {author} {\bibfnamefont
  {A.}~\bibnamefont {Péteut}}, \bibinfo {author} {\bibfnamefont
  {S.}~\bibnamefont {Hartman}}, \bibinfo {author} {\bibnamefont {rokvintar}},
  \bibinfo {author} {\bibfnamefont {R.}~\bibnamefont {Clarken}}, \bibinfo
  {author} {\bibfnamefont {B.}~\bibnamefont {Martins}}, \bibinfo {author}
  {\bibfnamefont {D.}~\bibnamefont {Allan}}, \bibinfo {author} {\bibfnamefont
  {T.}~\bibnamefont {Birke}}, \bibinfo {author} {\bibfnamefont {P.~R.}\
  \bibnamefont {Jemian}}, \bibinfo {author} {\bibfnamefont {N.}~\bibnamefont
  {Claesson}}, \bibinfo {author} {\bibfnamefont {J.}~\bibnamefont {Adelman}},
  \bibinfo {author} {\bibfnamefont {J.}~\bibnamefont {Dwyer}}, \bibinfo
  {author} {\bibfnamefont {H.}~\bibnamefont {Slepicka}}, \bibinfo {author}
  {\bibfnamefont {G.}~\bibnamefont {Brandl}}, \bibinfo {author} {\bibfnamefont
  {E.}~\bibnamefont {Greenberg}}, \bibinfo {author} {\bibfnamefont
  {D.}~\bibnamefont {Vine}},\ and\ \bibinfo {author} {\bibnamefont {André}},\
  }\href {https://doi.org/10.5281/zenodo.3241648} {\bibinfo {title}
  {pyepics/pyepics 3.4.0}} (\bibinfo {year} {2019})\BibitemShut {NoStop}%
\bibitem [{\citenamefont {Dalesio}\ \emph {et~al.}(1991)\citenamefont
  {Dalesio}, \citenamefont {Kraimer},\ and\ \citenamefont
  {Kozubal}}]{Dalesio1991}%
  \BibitemOpen
  \bibfield  {author} {\bibinfo {author} {\bibfnamefont {L.~R.}\ \bibnamefont
  {Dalesio}}, \bibinfo {author} {\bibfnamefont {M.~R.}\ \bibnamefont
  {Kraimer}},\ and\ \bibinfo {author} {\bibfnamefont {a.~J.}\ \bibnamefont
  {Kozubal}},\ }\bibfield  {title} {\bibinfo {title} {{EPICS Architecture}},\
  }\href@noop {} {\bibfield  {journal} {\bibinfo  {journal} {Proc. 1991
  ICALEPCS}\ } (\bibinfo {year} {1991})}\BibitemShut {NoStop}%
\bibitem [{\citenamefont {{GPy}}(2012)}]{gpy2014}%
  \BibitemOpen
  \bibfield  {author} {\bibinfo {author} {\bibnamefont {{GPy}}},\ }\href@noop
  {} {\bibinfo {title} {{GPy}: A gaussian process framework in python}},\
  \bibinfo {howpublished} {\url{http://github.com/SheffieldML/GPy}} (\bibinfo
  {year} {since 2012})\BibitemShut {NoStop}%
\bibitem [{\citenamefont {Virtanen}\ and\ \citenamefont {{SciPy 1.0
  Contributors}}(2020)}]{2020SciPy-NMeth}%
  \BibitemOpen
  \bibfield  {author} {\bibinfo {author} {\bibfnamefont {P.}~\bibnamefont
  {Virtanen}}\ and\ \bibinfo {author} {\bibnamefont {{SciPy 1.0
  Contributors}}},\ }\bibfield  {title} {\bibinfo {title} {{{SciPy} 1.0:
  Fundamental Algorithms for Scientific Computing in Python}},\ }\href
  {https://doi.org/10.1038/s41592-019-0686-2} {\bibfield  {journal} {\bibinfo
  {journal} {Nature Methods}\ }\textbf {\bibinfo {volume} {17}},\ \bibinfo
  {pages} {261} (\bibinfo {year} {2020})}\BibitemShut {NoStop}%
\bibitem [{\citenamefont {Terebilo}(2001)}]{Terebilo2001}%
  \BibitemOpen
  \bibfield  {author} {\bibinfo {author} {\bibfnamefont {A.}~\bibnamefont
  {Terebilo}},\ }\bibfield  {title} {\bibinfo {title} {{Accelerator toolbox for
  MATLAB}},\ }in\ \href@noop {} {\emph {\bibinfo {booktitle} {{Workshop on
  Performance Issues at Synchrotron Light Sources}}}}\ (\bibinfo {year}
  {2001})\BibitemShut {NoStop}%
\bibitem [{\citenamefont {Portmann}\ \emph {et~al.}(2005)\citenamefont
  {Portmann}, \citenamefont {Corbett},\ and\ \citenamefont
  {Terebilo}}]{Portmann2005}%
  \BibitemOpen
  \bibfield  {author} {\bibinfo {author} {\bibfnamefont {G.}~\bibnamefont
  {Portmann}}, \bibinfo {author} {\bibfnamefont {J.}~\bibnamefont {Corbett}},\
  and\ \bibinfo {author} {\bibfnamefont {A.}~\bibnamefont {Terebilo}},\
  }\bibfield  {title} {\bibinfo {title} {{An Accelerator control middle layer
  using Matlab}},\ }\href@noop {} {\bibfield  {journal} {\bibinfo  {journal}
  {Conf. Proc. C}\ }\textbf {\bibinfo {volume} {0505161}},\ \bibinfo {pages}
  {4009} (\bibinfo {year} {2005})}\BibitemShut {NoStop}%
\bibitem [{\citenamefont {Marsching}\ \emph {et~al.}(2011)\citenamefont
  {Marsching}, \citenamefont {Huttel}, \citenamefont {Klein},\ and\
  \citenamefont {Smale}}]{Marsching2011}%
  \BibitemOpen
  \bibfield  {author} {\bibinfo {author} {\bibfnamefont {S.}~\bibnamefont
  {Marsching}}, \bibinfo {author} {\bibfnamefont {E.}~\bibnamefont {Huttel}},
  \bibinfo {author} {\bibfnamefont {M.}~\bibnamefont {Klein}},\ and\ \bibinfo
  {author} {\bibfnamefont {N.~J.}\ \bibnamefont {Smale}},\ }\bibfield  {title}
  {\bibinfo {title} {{First Experience with the Matlab Middle Layer at ANKA}},\
  }in\ \href@noop {} {\emph {\bibinfo {booktitle} {ICALEPCS 2011}}}\ (\bibinfo
  {year} {2011})\BibitemShut {NoStop}%
\bibitem [{\citenamefont {Nelder}\ and\ \citenamefont
  {Mead}(1965)}]{Nelder1965}%
  \BibitemOpen
  \bibfield  {author} {\bibinfo {author} {\bibfnamefont {J.~A.}\ \bibnamefont
  {Nelder}}\ and\ \bibinfo {author} {\bibfnamefont {R.}~\bibnamefont {Mead}},\
  }\bibfield  {title} {\bibinfo {title} {A simplex method for function
  minimization},\ }\href {https://doi.org/10.1093/comjnl/7.4.308} {\bibfield
  {journal} {\bibinfo  {journal} {Comput. J.}\ }\textbf {\bibinfo {volume}
  {7}},\ \bibinfo {pages} {308} (\bibinfo {year} {1965})}\BibitemShut {NoStop}%
\bibitem [{\citenamefont {Frazier}(2018)}]{Frazier2018}%
  \BibitemOpen
  \bibfield  {author} {\bibinfo {author} {\bibfnamefont {P.~I.}\ \bibnamefont
  {Frazier}},\ }\href@noop {} {\bibinfo {title} {A tutorial on bayesian
  optimization}} (\bibinfo {year} {2018}),\ \Eprint
  {https://arxiv.org/abs/1807.02811} {arXiv:1807.02811 [stat.ML]} \BibitemShut
  {NoStop}%
\bibitem [{\citenamefont {St.~John}\ \emph {et~al.}(2021)\citenamefont
  {St.~John}, \citenamefont {Herwig}, \citenamefont {Kafkes}, \citenamefont
  {Mitrevski}, \citenamefont {Pellico}, \citenamefont {Perdue}, \citenamefont
  {Quintero-Parra}, \citenamefont {Schupbach}, \citenamefont {Seiya},
  \citenamefont {Tran}, \citenamefont {Schram}, \citenamefont {Duarte},
  \citenamefont {Huang},\ and\ \citenamefont {Keller}}]{StJohn2021}%
  \BibitemOpen
  \bibfield  {author} {\bibinfo {author} {\bibfnamefont {J.}~\bibnamefont
  {St.~John}}, \bibinfo {author} {\bibfnamefont {C.}~\bibnamefont {Herwig}},
  \bibinfo {author} {\bibfnamefont {D.}~\bibnamefont {Kafkes}}, \bibinfo
  {author} {\bibfnamefont {J.}~\bibnamefont {Mitrevski}}, \bibinfo {author}
  {\bibfnamefont {W.~A.}\ \bibnamefont {Pellico}}, \bibinfo {author}
  {\bibfnamefont {G.~N.}\ \bibnamefont {Perdue}}, \bibinfo {author}
  {\bibfnamefont {A.}~\bibnamefont {Quintero-Parra}}, \bibinfo {author}
  {\bibfnamefont {B.~A.}\ \bibnamefont {Schupbach}}, \bibinfo {author}
  {\bibfnamefont {K.}~\bibnamefont {Seiya}}, \bibinfo {author} {\bibfnamefont
  {N.}~\bibnamefont {Tran}}, \bibinfo {author} {\bibfnamefont {M.}~\bibnamefont
  {Schram}}, \bibinfo {author} {\bibfnamefont {J.~M.}\ \bibnamefont {Duarte}},
  \bibinfo {author} {\bibfnamefont {Y.}~\bibnamefont {Huang}},\ and\ \bibinfo
  {author} {\bibfnamefont {R.}~\bibnamefont {Keller}},\ }\bibfield  {title}
  {\bibinfo {title} {Real-time artificial intelligence for accelerator control:
  A study at the fermilab booster},\ }\href
  {https://doi.org/10.1103/PhysRevAccelBeams.24.104601} {\bibfield  {journal}
  {\bibinfo  {journal} {Phys. Rev. Accel. Beams}\ }\textbf {\bibinfo {volume}
  {24}},\ \bibinfo {pages} {104601} (\bibinfo {year} {2021})}\BibitemShut
  {NoStop}%
\bibitem [{\citenamefont {Eichler}\ \emph {et~al.}(2021)\citenamefont
  {Eichler}, \citenamefont {Bründermann}, \citenamefont {Burkart},
  \citenamefont {Kaiser}, \citenamefont {Kuropka}, \citenamefont {{Santamaria
  Garcia}}, \citenamefont {Stein},\ and\ \citenamefont {Xu}}]{Eichler2021}%
  \BibitemOpen
  \bibfield  {author} {\bibinfo {author} {\bibfnamefont {A.}~\bibnamefont
  {Eichler}}, \bibinfo {author} {\bibfnamefont {E.}~\bibnamefont
  {Bründermann}}, \bibinfo {author} {\bibfnamefont {F.}~\bibnamefont
  {Burkart}}, \bibinfo {author} {\bibfnamefont {J.}~\bibnamefont {Kaiser}},
  \bibinfo {author} {\bibfnamefont {W.}~\bibnamefont {Kuropka}}, \bibinfo
  {author} {\bibfnamefont {A.}~\bibnamefont {{Santamaria Garcia}}}, \bibinfo
  {author} {\bibfnamefont {O.}~\bibnamefont {Stein}},\ and\ \bibinfo {author}
  {\bibfnamefont {C.}~\bibnamefont {Xu}},\ }\bibfield  {title} {\bibinfo
  {title} {{First Steps Toward an Autonomous Accelerator, a Common Project
  Between DESY and KIT}},\ }in\ \href
  {https://doi.org/10.18429/JACoW-IPAC2021-TUPAB298} {\emph {\bibinfo
  {booktitle} {Proc. IPAC'21}}},\ \bibinfo {series and number} {\bibinfo
  {series} {International Particle Accelerator Conference}\ No.~\bibinfo
  {number} {12}}\ (\bibinfo  {publisher} {JACoW Publishing, Geneva,
  Switzerland},\ \bibinfo {year} {2021})\ pp.\ \bibinfo {pages} {2182--2185},\
  \bibinfo {note}
  {https://doi.org/10.18429/JACoW-IPAC2021-TUPAB298}\BibitemShut {NoStop}%
\bibitem [{\citenamefont {Boltz}\ \emph {et~al.}(2020)\citenamefont {Boltz},
  \citenamefont {Bründermann}, \citenamefont {Caselle}, \citenamefont
  {Kopmann}, \citenamefont {Mexner}, \citenamefont {Müller},\ and\
  \citenamefont {Wang}}]{Boltz2019}%
  \BibitemOpen
  \bibfield  {author} {\bibinfo {author} {\bibfnamefont {T.}~\bibnamefont
  {Boltz}}, \bibinfo {author} {\bibfnamefont {E.}~\bibnamefont {Bründermann}},
  \bibinfo {author} {\bibfnamefont {M.}~\bibnamefont {Caselle}}, \bibinfo
  {author} {\bibfnamefont {A.}~\bibnamefont {Kopmann}}, \bibinfo {author}
  {\bibfnamefont {W.}~\bibnamefont {Mexner}}, \bibinfo {author} {\bibfnamefont
  {A.-S.}\ \bibnamefont {Müller}},\ and\ \bibinfo {author} {\bibfnamefont
  {W.}~\bibnamefont {Wang}},\ }\bibfield  {title} {\bibinfo {title}
  {{Accelerating Machine Learning for Machine Physics (an AMALEA-project at
  KIT)}},\ }in\ \href {https://doi.org/10.18429/JACoW-ICALEPCS2019-TUCPL06}
  {\emph {\bibinfo {booktitle} {Proc. ICALEPCS'19}}},\ \bibinfo {series and
  number} {\bibinfo {series} {International Conference on Accelerator and Large
  Experimental Physics Control Systems}\ No.~\bibinfo {number} {17}}\ (\bibinfo
   {publisher} {JACoW Publishing, Geneva, Switzerland},\ \bibinfo {year}
  {2020})\ pp.\ \bibinfo {pages} {781--788},\ \bibinfo {note}
  {https://doi.org/10.18429/JACoW-ICALEPCS2019-TUCPL06}\BibitemShut {NoStop}%
\bibitem [{\citenamefont {Wang}\ \emph {et~al.}(2021)\citenamefont {Wang},
  \citenamefont {Caselle}, \citenamefont {Boltz}, \citenamefont {Blomley},
  \citenamefont {Brosi}, \citenamefont {Dritschler}, \citenamefont {Ebersoldt},
  \citenamefont {Kopmann}, \citenamefont {Santamaria~Garcia}, \citenamefont
  {Schreiber}, \citenamefont {Bründermann}, \citenamefont {Weber},
  \citenamefont {Müller},\ and\ \citenamefont {Fang}}]{Wang2021}%
  \BibitemOpen
  \bibfield  {author} {\bibinfo {author} {\bibfnamefont {W.}~\bibnamefont
  {Wang}}, \bibinfo {author} {\bibfnamefont {M.}~\bibnamefont {Caselle}},
  \bibinfo {author} {\bibfnamefont {T.}~\bibnamefont {Boltz}}, \bibinfo
  {author} {\bibfnamefont {E.}~\bibnamefont {Blomley}}, \bibinfo {author}
  {\bibfnamefont {M.}~\bibnamefont {Brosi}}, \bibinfo {author} {\bibfnamefont
  {T.}~\bibnamefont {Dritschler}}, \bibinfo {author} {\bibfnamefont
  {A.}~\bibnamefont {Ebersoldt}}, \bibinfo {author} {\bibfnamefont
  {A.}~\bibnamefont {Kopmann}}, \bibinfo {author} {\bibfnamefont
  {A.}~\bibnamefont {Santamaria~Garcia}}, \bibinfo {author} {\bibfnamefont
  {P.}~\bibnamefont {Schreiber}}, \bibinfo {author} {\bibfnamefont
  {E.}~\bibnamefont {Bründermann}}, \bibinfo {author} {\bibfnamefont
  {M.}~\bibnamefont {Weber}}, \bibinfo {author} {\bibfnamefont {A.-S.}\
  \bibnamefont {Müller}},\ and\ \bibinfo {author} {\bibfnamefont
  {Y.}~\bibnamefont {Fang}},\ }\bibfield  {title} {\bibinfo {title}
  {Accelerated deep reinforcement learning for fast feedback of beam dynamics
  at kara},\ }\href {https://doi.org/10.1109/TNS.2021.3084515} {\bibfield
  {journal} {\bibinfo  {journal} {IEEE Transactions on Nuclear Science}\
  }\textbf {\bibinfo {volume} {68}},\ \bibinfo {pages} {1794} (\bibinfo {year}
  {2021})}\BibitemShut {NoStop}%
\end{thebibliography}%

\end{document}